\documentclass[journal,onecolumn]{IEEEtran}
\usepackage{amsmath,amsfonts,amssymb}
\usepackage{multirow}
\usepackage{booktabs}
\usepackage{algorithmic}
\usepackage{algorithm}
\usepackage{abstract}
\usepackage{array}
\usepackage[justification=centering,caption=false,font=normalsize,labelfont=sf,textfont=sf]{subfig}
\usepackage{textcomp}
\usepackage{stfloats}
\usepackage{url}
\usepackage{bbm}
\usepackage{comment}
\usepackage{verbatim}
\usepackage{graphicx}
\usepackage[noadjust]{cite}
\usepackage{xcolor}
\usepackage{tikz}
\usetikzlibrary{matrix}
\usepackage{enumerate}

\newtheorem{theorem}{Theorem}
\newtheorem{definition}{Definition}
\newtheorem{lemma}{Lemma}
\newtheorem{construction}{Construction}
\newtheorem{remark}{Remark}

\newtheorem{corollary}{Corollary}

\begin{document}

\title{Bounds and Constructions of $\ell$-Read Codes under the Hamming Metric}

\author{Yubo~Sun and Gennian~Ge%
\thanks{The research of G. Ge was supported by the National Key Research and Development Program of China under Grant 2020YFA0712100, the National Natural Science Foundation of China under Grant 12231014, and Beijing Scholars Program.}
\thanks{Y. Sun ({\tt 2200502135@cnu.edu.cn}) and G. Ge ({\tt gnge@zju.edu.cn}) are with the School of Mathematical Sciences, Capital Normal University, Beijing 100048, China. ({\em Corresponding author:
Gennian Ge.})}
}


\maketitle

\begin{abstract}
Nanopore sequencing is a promising technology for DNA sequencing.
In this paper, we investigate a specific model of the nanopore sequencer, which takes a $q$-ary sequence of length $n$ as input and outputs a vector of length $n+\ell-1$ referred to as an $\ell$-read vector where the $i$-th entry is a multi-set composed of the $\ell$ elements located between the $(i-\ell+1)$-th and $i$-th positions of the input sequence. 
Considering the presence of substitution errors in the output vector, we study $\ell$-read codes under the Hamming metric.
An $\ell$-read $(n,d)_q$-code is a set of $q$-ary sequences of length $n$ in which the Hamming distance between $\ell$-read vectors of any two distinct sequences is at least $d$. 
We first improve the result of Banerjee \emph{et al.}, who studied $\ell$-read $(n,d)_q$-codes with the constraint $\ell\geq 3$ and $d=3$.
Then, we investigate the bounds and constructions of $2$-read codes with a minimum distance of $3$, $4$, and $5$, respectively.
Our results indicate that when $d \in \{3,4\}$, the optimal redundancy of $2$-read $(n,d)_q$-codes is $o(\log_q n)$, while for $d=5$ it is $\log_q n+o(\log_q n)$.
Additionally, we establish an equivalence between $2$-read $(n,3)_q$-codes and classical $q$-ary single-insertion reconstruction codes using two noisy reads.
We improve the lower bound on the redundancy of classical $q$-ary single-insertion reconstruction codes as well as the upper bound on the redundancy of classical $q$-ary single-deletion reconstruction codes when using two noisy reads.
Finally, we study $\ell$-read codes under the reconstruction model.
\end{abstract}



\section{Introduction}

\IEEEPARstart{T}{he} exponential growth of digital data in the era of big data poses unprecedented challenges to the storage capacities of existing storage media \cite{Rydning-22-TR-background}.
DNA provides a promising solution to these challenges due to its ultra-high storage density and unmatched longevity \cite{DNA-21}.
However, the high cost of sequencing technologies has hindered the progress of DNA storage.
Nanopore sequencing \cite{Deamer-16-NB-nanopore,Kasianowicz-96-NB-nanopore,Laszlo-14-NB-nanopore} seems to offer a pathway towards making large-scale DNA storage economically viable, thanks to its support for long reads, low cost, and better portability.
The nanopore sequencer operates by passing a DNA fragment through a  tiny pore at a relatively stable speed within a membrane.
As the DNA moves through the pore, a constant number of nucleotides occupy the pore at any given moment, and the presence of specific nucleotides in the nanopore influences the ionic current flowing through it. The sequencer captures the electrical signals produced during this process, enabling the prediction of the nucleotide sequence within the DNA fragment.
Despite the efficiency of nanopore sequencing in various respects, certain physical aspects of the process have the potential to distort the final readout. 
Factors such as the simultaneous presence of multiple nucleotides in the pore, irregular DNA movement, and random noise in ionic current measurements can lead to errors like inter-symbol interference, duplications, deletions, and substitutions in the sequencing results. Previous works in this field \cite{Hulett-21-ISIT-nanopore, Mao-18-IT-nanopore, McBain-22-ISIT-nanopore, McBain-23-ISIT-nanopore, Banerjee-23-arxiv-nanopore, Banerjee-24-arxiv-nanopore, Yerushalmi-24-arxiv-nanopore, Vidal-23-ICC-nanopore, Vidal-23-ISIT-nanopore} have focused on developing accurate mathematical models for the sequencer and designing error-correcting codes that leverage these models to efficiently rectify errors in the readouts.

In this paper, we investigate a specific model introduced in \cite{Banerjee-23-arxiv-nanopore}, which drew inspiration from the work in \cite{Mao-18-IT-nanopore} and shares similarities with the transverse-read channel described in \cite{Chee-23-JSAIT-nanopore}.
This model takes a $q$-ary sequence of length $n$ as input and outputs a vector of length $n+\ell-1$, known as an $\ell$-read vector. 
The $i$-th entry of the output vector is a multi-set containing the $\ell$ elements positioned between the $(i-\ell+1)$-th and $i$-th positions of the input sequence.
Given the potential for random noise to introduce substitution errors in ionic current measurements, we focus on designing error-correcting codes that can rectify substitution errors within the framework of this model.

A set of $q$-ary sequences of length $n$ is referred to as an $\ell$-read $(n,d)_q$-code, if the Hamming distance between $\ell$-read vectors of any two distinct sequences in the set is at least $d$. 
While previous work \cite{Banerjee-23-arxiv-nanopore} has focused on $\ell$-read codes, where $\ell \geq 3$, with a minimum Hamming distance of three, higher error rates necessitate considering larger minimum distance. 
In this paper, we take the initial step towards exploring the bounds and constructions of $\ell$-read codes with a larger minimum distance.
Specifically, we primarily consider the scenario where $\ell=2$ and $d\in \{3,4,5\}$.

Note that in $\ell$-read codes, the focus is on the Hamming distance between the $\ell$-read vectors of two sequences, rather than the Hamming distance between the two sequences themselves. 
In Section \ref{sec:condition}, we characterize the structure of two sequences when their $2$-read vectors have a Hamming distance of exactly $d$. 
By understanding when $2$-read vectors have a fixed Hamming distance, in Sections \ref{sec:d=3}, \ref{sec:d=4}, and \ref{sec:d=5}, we delve into the bounds and constructions of $2$-read codes with a minimum distance of $3$, $4$, and $5$, respectively.
In Section \ref{sec:d=3}, we extend our investigation to the general read length $\ell \geq 2$ rather than restricting to $\ell=2$. 
We find the difference between the cases where $\ell \geq 3$ and $\ell= 2$.
More specifically, our results show that for $\ell \geq 3$, the optimal redundancy of $\ell$-read $(n,3)_q$-codes asymptotically approaches to $\log_q\log_q n-\log_q 2$, which improves the work by Banerjee \emph{et al.} \cite{Banerjee-23-arxiv-nanopore}, while for $\ell=2$ it ranges from $\log_q\log_q n-o(1)$ to $\log_q\log_q n+ 1-\log_q 2+o(1)$. 
Additionally, we establish an equivalence between $2$-read $(n,3)_q$-codes and classical $(n,2;\mathcal{I}_1)_q$-reconstruction codes (to be defined later in Definition
\ref{def:recon}).
We improve the lower bound on the redundancy of classical $(n,2;\mathcal{I}_1)_q$-reconstruction codes as well as the upper bound on the redundancy of classical $(n,2;\mathcal{D}_1)_q$-reconstruction codes (to be defined later in Definition
\ref{def:recon}).
Particularly, when $q=2$, we determine the optimal values of the asymptotic redundancy of classical $(n,2;\mathcal{I}_1)_q$-reconstruction codes and classical $(n,2;\mathcal{D}_1)_q$-reconstruction codes, respectively, and derive an unexpected result that the optimal redundancy of classical $(n,2;\mathcal{D}_1)_2$-reconstruction codes is less than that of classical $(n,2;\mathcal{I}_1)_2$-reconstruction codes.
In Section \ref{sec:d=4}, we focus on the constructions of $2$-read $(n,4)_q$-codes, as the lower bound can be derived from that of $2$-read $(n,3)_q$-codes directly.
We design $2$-read $(n,4)_q$-codes with $2\log_q\log_q n+ \log_q(3(q-1)+1)+2\log_q 4+o(1)$ and $\log_2\log_2 n+\log_2 6+o(1)$ bits of redundancy for $q\geq 3$ and $q=2$, respectively.
In Section \ref{sec:d=5}, we establish a lower bound on the redundancy of $2$-read $(n,5)_q$-codes by linking them to classical $(n,3)_q$-codes and then focus on the construction of $2$-read $(n,5)_q$-codes.
Our results show that the optimal redundancy of $2$-read $(n,5)_q$-codes ranges from $\log_q n+O(1)$ to $\log_q n+ 4\log_q\log_q n+O(1)$. 
In Section \ref{sec:recon}, we study $\ell$-read codes under the reconstruction model and prove that the maximum intersection size between two $\ell$-read vectors with a Hamming distance of $d$, obtained by two $q$-ary sequences of length $n$, is no more than the maximum intersection size between two $\binom{q+\ell-1}{\ell}$-ary sequences of length $(n+\ell-1)$, also with a Hamming distance of $d$.
Finally, Section \ref{sec:concl} concludes the paper.
To facilitate comparison with prior works, we summarize our main contributions in Tables \ref{tab:result} and \ref{tab:result'}.

\begin{table}
\renewcommand\arraystretch{1.5}
    \caption{Lower and upper bounds on the redundancy of optimal $\ell$-read $(n,d)_q$-codes. Terms of order $o(1)$ are omitted.}
    \centering
    \begin{tabular}{ccccc}
        \hline
        \hline
         & Parameters $(\ell,d,q)$ & Lower Bound & Upper Bound & Remark\\
        \hline
        Previous Results & $\ell\geq 3, d=3, q\geq 2$ & $\log_q\log_q n-\log_q2$ & $\log_q\log_q n+\log_q (2q(q-1))$ & Theorems 5 and 6 of \cite{Banerjee-23-arxiv-nanopore}\\
        \hline
        \multirow{4}{*}{Our Results} & $\ell\geq 3, d=3, q\geq 2$ &  & $\log_q\log_q n-\log_q2$ & Theorem \ref{thm:code}\\
        & $\ell=2, d=3, q\geq 2$ & $\log_q\log_q n$ & $\log_q\log_q n+1-\log_q 2$ & Corollary \ref{cor:d=3,l=2}\\
        & $\ell=2, d=4, q\geq 3$ & $\log_q\log_q n$ & $2\log_q\log_q n+ O(1)$ & Corollary \ref{cor:d=4,l=2,q>=3}\\
        & $\ell=2, d=4, q= 2$ & $\log_2\log_2 n$ & $\log_2\log_2 n+\log_2 6$ & Corollary \ref{cor:d=4,l=2,q=2}\\
        & $\ell=2, d=5, q\geq 2$ & $\log_q n+O(1)$ & $\log_q n+4\log_q\log_q n+O(1)$ & Corollary \ref{cor:d=5,l=2}\\
        \hline
        \hline
    \end{tabular}
    \label{tab:result}
\end{table}

\begin{table}
\renewcommand\arraystretch{1.5}
    \caption{Lower and upper bounds on the redundancy of optimal classical $(n,2;\mathcal{B})$-reconstruction codes. Terms of order $o(1)$ are omitted.}
    \centering
    \begin{tabular}{ccccc}
        \hline
        \hline
         & Error Ball $\mathcal{B}$ & Lower Bound & Upper Bound & Remark \\
        \hline
        \multirow{1}{*}{Previous Results} & $\mathcal{B}\in \{\mathcal{I}_1,\mathcal{D}_1\}$ & $\log_q\log_q n -\log_q 2$ & $\log_q\log_q n +1$ & Proposition 10 of \cite{Chrisnata-22-IT-reconstr-del} and Corollary 18 of \cite{Cai-22-IT-recon-edit}\\
        \hline
        \multirow{2}{*}{Our Results} & $\mathcal{B}= \mathcal{I}_1$ & $\log_q\log_q n$ & $\log_q\log_q n +1-\log_q 2$ & Corollary \ref{cor:d=3,l=2}\\
        & $\mathcal{B}= \mathcal{D}_1$ & & $\log_q\log_q n +\min\{\log_q(q-1)/2,0\}$ & Corollary \ref{cor:del-recon} \\
        \hline
        \hline
    \end{tabular}
    \label{tab:result'}
\end{table}

\section{Preliminaries}

\subsection{Notations}
We introduce the following notations to be used throughout the paper.

Let $\Sigma_q$ denote the alphabet set $\{0,1,\ldots,q-1\}$ and $\Sigma_q^n$ denote the set of all sequences of length $n$ over the alphabet set $\Sigma_q$, where $q \geq 2$ and $n \geq 0$ are integers.
Here, $\Sigma_q^0$ is referred to as the empty set $\emptyset$.
Additionally, let $\Sigma_q^{\geq n} \triangleq \cup_{i=n}^{\infty} \Sigma_i$ be the set of all sequences of length at least $n$ over $\Sigma_q$.
Let $|\boldsymbol{x}|$ be the \emph{length} of $\boldsymbol{x}$ when $\boldsymbol{x}$ is a sequence, and $|\mathcal{S}|$ be the \emph{size} of $\mathcal{S}$ when $\mathcal{S}$ is a set.

For two integers $i$ and $j$ such that $i \leq j$, let $[i,j]$ denote the set $\{i,i+1,\ldots,j\}$.
Furthermore, when $i>j$, $[i,j]$ is referred to as the empty set $\emptyset$.
For any sequence $\boldsymbol{x} \in \Sigma_q^n$, $x[i]$ is referred to as its $i$-th entry for $i \in [1,n]$. 
Additionally, we set $x[i]=0$ when $i \not\in [1,n]$. 
Then the sequence $\boldsymbol{x}$ can be written as either $x[1]x[2]\cdots x[n]$ or $(x[1],x[2],\ldots,x[n])$.
If $\boldsymbol{y}\in \Sigma_q^m$ is also a $q$-ary sequence, let $(\boldsymbol{x},\boldsymbol{y})$ or $\boldsymbol{xy}$ denote the \emph{concatenation} of $\boldsymbol{x}$ and $\boldsymbol{y}$,
Moreover, let $\boldsymbol{x}^n$ denote the concatenation of $n$ copies of $\boldsymbol{x}$.
If there exist two distinct symbols $a,b \in \Sigma_q$ such that
\begin{equation*}
    \boldsymbol{x}= 
    \begin{cases}
      (ab)^{n/2}, & \mbox{if } 2|n \\
      (ab)^{(n-1)/2}a, & \mbox{otherwise}
    \end{cases},
\end{equation*}
we say that $\boldsymbol{x}$ is an \emph{alternating sequence}, and use $\boldsymbol{\alpha}_n(ab)$ to denote $\boldsymbol{x}$.
Note that $\boldsymbol{\alpha}_n(ab)=a$ when $n=1$ and $\boldsymbol{\alpha}_n(ab)=\emptyset$ when $n=0$.
If $n \geq m$, we say that $\boldsymbol{y}$ is a \emph{substring} of $\boldsymbol{x}$ if $\boldsymbol{y}= \boldsymbol{x}[i,j] \triangleq (x[i], x[i+1], \ldots, x[j])$ for some $i,j \in [1,n]$.
Moreover, we say that $\boldsymbol{x}[i,j]$ is an \emph{alternating substring} of $\boldsymbol{x}$ if it is an alternating sequence.
Let $\mathcal{ALL}(n,P)$ be the set of all length-$n$ sequences over $\Sigma_q$ in which each alternating substring is of length at most $P$.

The \emph{inversion number} of $\boldsymbol{x} \in \Sigma_q^n$ is defined as $\mathrm{Inv} (\boldsymbol{x}) \triangleq \left| \{(i,j) : 1 \leq i < j \leq n, x[i] > x[j] \} \right|$.
Let $k$ be a non-negative integer, the \emph{$k$-th order VT syndrome} of $\boldsymbol{x}$ is defined as $\mathrm{VT}^{(k)}(\boldsymbol{x})=\sum_{i=1}^n i^k x[i]$.

\subsection{Models and Codes}

We begin with the definition of $\ell$-read vectors.

\begin{definition}
  The \emph{composition} of a sequence $\boldsymbol{x}\in \Sigma_q^n$ is defined as the multi-set $c(\boldsymbol{x})\triangleq \{\{x_1,x_2,\ldots,x_n\}\}$.
  For positive integers $\ell$ and $n$, the \emph{$\ell$-read vector} of $\boldsymbol{x}$ is of length $n+\ell-1$ and is defined as 
  \begin{equation*}
    \mathcal{R}_{\ell}(\boldsymbol{x})= (c(\boldsymbol{x}[2-\ell,1]), c(\boldsymbol{x}[3-\ell,2]), \ldots, c(\boldsymbol{x}[n,n+\ell-1])).
  \end{equation*}
\end{definition}

Note that when $\ell=1$, we have $\mathcal{R}_{\ell}(\boldsymbol{x})= \boldsymbol{x}$ by setting $\{x[i]\}\triangleq x[i]$ for $i \in [1,n]$.
Therefore, in the rest of this paper, we only consider the case of $\ell \geq 2$.
Moreover, when $\ell=2$, we abbreviate $\mathcal{R}_{\ell}(\boldsymbol{x})$ as $\mathcal{R}(\boldsymbol{x})$.

\begin{remark}
     The above definition of an $\ell$-read vector is essentially the same as that which was introduced in \cite{Banerjee-23-arxiv-nanopore}.
     Note that $x[i]$ can be determined by $c(\boldsymbol{x}[i-\ell+1,i])$ and $x[j]$ for $j \in [i-\ell+1,i-1]$, it follows by $x[i]=0$ for $i \not\in [1,n]$ that there is a one to one correspondence between $\boldsymbol{x}$ and $\mathcal{R}_{\ell}(\boldsymbol{x})$.
\end{remark}

We proceed to define the error-correcting codes and reconstruction codes that will be investigated for the remainder of this paper.

\begin{definition}
Given two sequences or vectors $\boldsymbol{x}$ and $\boldsymbol{y}$ of the same length, the \emph{Hamming distance} between them, denoted as $d_H(\boldsymbol{x},\boldsymbol{y})$, is defined as the number of positions where the corresponding entries differ.
\end{definition}

\begin{definition}
   For $d \geq 1$, a set $\mathcal{C} \in \Sigma_q^n$ is referred to as an \emph{$\ell$-read $(n,d)_q$-code} or an $\ell$-read code with a minimum distance of $d$ if for any two distinct sequences $\boldsymbol{x}, \boldsymbol{y} \in \mathcal{C}$, it holds that $d_H(\mathcal{R}_{\ell}(\boldsymbol{x}),\mathcal{R}_{\ell}(\boldsymbol{y}))\geq d$.
   Moreover, $\mathcal{C} \in \Sigma_q^n$ is referred to as a \emph{classical $(n,d)_q$-code} or a classical code with a minimum distance of $d$ if for any two distinct sequences $\boldsymbol{x}, \boldsymbol{y} \in \mathcal{C}$, it holds that $d_H(\boldsymbol{x},\boldsymbol{y})\geq d$. 
\end{definition}

\begin{definition}
Given a positive integer $t$ and a sequence $\boldsymbol{x}$, let $\mathcal{S}_t(\boldsymbol{x})$ be the \emph{$t$-substitution ball} of $\boldsymbol{x}$, which is defined as the set of all sequences that can be obtained by substituting at most $t$ symbols from $\boldsymbol{x}$.
Moreover, let $\mathcal{D}_t(\boldsymbol{x})$ (or $\mathcal{I}_t(\boldsymbol{x})$) be the \emph{$t$-deletion ball} (or \emph{$t$-insertion ball}) of $\boldsymbol{x}$, which is defined as the set of all sequences that can be obtained by deleting (or inserting, respectively) exactly $t$ symbols from $\boldsymbol{x}$.
\end{definition}

\begin{definition}\label{def:recon}
   For $t \geq 1$ and $N \geq 1$, a set $\mathcal{C} \in \Sigma_q^n$ is referred to as an \emph{$\ell$-read $(n,N;\mathcal{S}_t)_q$-reconstruction code} if for any two distinct sequences $\boldsymbol{x}, \boldsymbol{y} \in \mathcal{C}$, it holds that $|\mathcal{S}_t(\mathcal{R}_{\ell}(\boldsymbol{x})) \cap \mathcal{S}_t(\mathcal{R}_{\ell}(\boldsymbol{y}))|\leq N-1$.
   Moreover, let $\mathcal{B} \in \{\mathcal{S}_t,\mathcal{D}_t,\mathcal{I}_t\}$, we say that $\mathcal{C}$ is a \emph{classical $(n,N;\mathcal{B})_q$-reconstruction code} if for any two distinct sequences $\boldsymbol{x}, \boldsymbol{y} \in \mathcal{C}$, it holds that $|\mathcal{B}(\boldsymbol{x}) \cap \mathcal{B}(\boldsymbol{y})|\leq N-1$.
\end{definition}

To evaluate an error-correcting code or a reconstruction code, we calculate its \emph{redundancy}, which is defined as $r(\mathcal{C})\triangleq n- \log_q |\mathcal{C}|$.

\subsection{Previous Results}

In this subsection, we present several known results related to $\ell$-read $(n,3)_q$-codes and classical $(n,2;\mathcal{B})_q$-reconstruction codes, where $\mathcal{B} \in \{\mathcal{D}_1,\mathcal{I}_1\}$, which will be used later.

\subsubsection{$\ell$-Read $(n,3)_q$-Codes with $\ell \geq 3$}

\begin{lemma}[Theorem 2 of \cite{Banerjee-23-arxiv-nanopore}]\label{lem:1sub}
  Assume $\ell \geq 3$ and $\boldsymbol{x}\neq \boldsymbol{y}\in \Sigma_q^n$,  $d_H(\mathcal{R}_{\ell}(\boldsymbol{x}),\mathcal{R}_{\ell}(\boldsymbol{y})) \leq 2$ holds if and only if there exist $t+2$ sequences $\boldsymbol{u}, \boldsymbol{w} \in \Sigma_q^{\geq 0}, \boldsymbol{v}_1, \ldots, \boldsymbol{v}_t \in \Sigma_q^{\ell-2}$ for some $t\geq 0$, and two distinct symbols $a, b \in \Sigma_q$, such that
  \begin{equation*}
    \begin{cases}
      \boldsymbol{x}= (\boldsymbol{u}, (a, b, \boldsymbol{v}_1), (a,b, \boldsymbol{v}_2), \ldots, (a,b, \boldsymbol{v}_t), a,b, \boldsymbol{w}); \\
      \boldsymbol{y}= (\boldsymbol{u}, (b, a, \boldsymbol{v}_1), (b,a, \boldsymbol{v}_2), \ldots, (b,a, \boldsymbol{v}_t), b,a, \boldsymbol{w}).
    \end{cases}
  \end{equation*} 
\end{lemma}

\begin{lemma}[Theorems 5 and 6 of \cite{Banerjee-23-arxiv-nanopore}]\label{lem:l>=3}
  Assume $\ell \geq 3$, the optimal redundancy of $\ell$-read $(n,3)_q$-codes is lower bounded by $\log_q\log_q n- \log_q \binom{q}{2}- o(1)$ (there is a flaw in their calculation of this value, and the correct value is $\log_q \log_q n- \log_q 2- o(1)$), and it is upper bounded by $\log_q \log_q n+ \log_q 2(q-1)+1+o(1)$.
\end{lemma}

\subsubsection{$(n,2;\mathcal{B})_q$-Reconstruction Codes where $\mathcal{B} \in \{\mathcal{D}_1,\mathcal{I}_1\}$}

\begin{lemma}[Definition 8 and Proposition 9 of \cite{Cai-22-IT-recon-edit}]\label{lem:indel}
  For two distinct sequences $\boldsymbol{x}, \boldsymbol{y} \in \Sigma_q^n$, the following statements are true.
  \begin{itemize}
    \item $|\mathcal{D}_1(\boldsymbol{x}) \cap \mathcal{D}_1(\boldsymbol{y})| =2$ holds if and only if there exist two sequences $\boldsymbol{u}, \boldsymbol{v}\in \Sigma_q^{\geq 0}$, two distinct symbols $a,b \in \Sigma_q$, and an integer $t \geq 2$, such that 
        \begin{equation*}
            \begin{cases}
              \boldsymbol{x}= (\boldsymbol{u}, \boldsymbol{\alpha}_t(ab), \boldsymbol{v}); \\
              \boldsymbol{y}= (\boldsymbol{u}, \boldsymbol{\alpha}_t(ba), \boldsymbol{v}).
            \end{cases}
          \end{equation*}
    \item $|\mathcal{I}_1(\boldsymbol{x}) \cap \mathcal{I}_1(\boldsymbol{y})| =2$ holds if and only if there exist two sequences $\boldsymbol{u}, \boldsymbol{v}\in \Sigma_q^{\geq 0}$, two distinct symbols $a,b \in \Sigma_q$, and an integer $t \geq 1$, such that 
        \begin{equation*}
            \begin{cases}
              \boldsymbol{x}= (\boldsymbol{u}, \boldsymbol{\alpha}_t(ab), \boldsymbol{v}); \\
              \boldsymbol{y}= (\boldsymbol{u}, \boldsymbol{\alpha}_t(ba), \boldsymbol{v}).
            \end{cases}
          \end{equation*}
  \end{itemize}
\end{lemma}

\begin{lemma}[Proposition 10 of \cite{Chrisnata-22-IT-reconstr-del} or Theorem 5.4 of \cite{Sun-23-IT-BDR}]\label{lem:ins_bound}
  Let $\mathcal{B} \in \{\mathcal{D}_1,\mathcal{I}_1\}$, the optimal redundancy of classical $(n,2;\mathcal{B})_q$-reconstruction codes is lower bounded by $\log_q \log_q n- \log_q 2- o(1)$.
\end{lemma}

\begin{lemma}[Lemma 6.10 of \cite{Sun-23-IT-BDR}]
Assume $P \geq \lceil \log_q n \rceil + 3$, we have $|\mathcal{ALL}(n,P)| \geq \frac{q^n}{2}$.
\end{lemma}

\begin{lemma}[Theorem 17 and Corollary 18 of \cite{Cai-22-IT-recon-edit}]\label{lem:ins-recon}
    Assume $P\geq \lceil \log_q n \rceil+3$ is an even integer.
    Let $a_1 \in [0,q-1]$ and $a_2 \in [0,\frac{P}{2}]$, the code
    \begin{align*}
        \mathcal{C}_P(a_1,a_2) \triangleq \{ \boldsymbol{x}
        & \in \mathcal{ALL}(n,P):~\mathrm{VT}^{(0)}(\boldsymbol{x}) \equiv a_1 \pmod{q},~{\rm Inv} (\boldsymbol{x}) \equiv a_2 \pmod{1+P/2} \},
    \end{align*}
    is a classical $(n,2;\mathcal{B})_q$-reconstruction code, where $\mathcal{B} \in \{\mathcal{D}_1,\mathcal{I}_1\}$.
     Moreover, if $P$ is the smallest integer larger than $\lceil \log_q n \rceil+3$, there exists a choice of $a_1$ and $a_2$ such that $r(\mathcal{C}_P(a_1,a_2)) \leq \log_q \log_q n+ 1+ o(1)$.
\end{lemma}

Theoretically, in Lemma \ref{lem:ins-recon}, we can slightly reduce the code redundancy by setting $P\geq \log_q n + \log_q \log_q n$.

\begin{lemma}\label{lem:all}
   Assume $n \rightarrow \infty$. Let $P$ be the smallest even integer larger than $\log_q n + \log_q \log_q n$ and let $\mathcal{C}_P(a_1,a_2)$ be the code constructed in Lemma \ref{lem:ins-recon}, there exists a choice of $a_1$ and $a_2$ such that $r(\mathcal{C}_P(a_1,a_2)) \leq \log_q \log_q n+ 1-\log_q 2+ o(1)$.
\end{lemma}

\begin{IEEEproof}
    We first show that $|\mathcal{ALL}(n,P)|= (1-o(1))q^n$. 
    Choose a sequence $\boldsymbol{x}\in \Sigma_q^n$ uniformly at random, the probability that $\boldsymbol{x} \in \mathcal{ALL}(n,P)$ can be calculated as
    \begin{align*}
    \mathbbm{P} (\boldsymbol{x} \in \mathcal{ALL}(n,P))
    & \geq 1- n \frac{q(q-1)}{q^P}
    \geq 1- \frac{nq(q-1)}{n\log_q n}= 1-o(1).
    \end{align*}
    It follows that $|\mathcal{ALL}(n,P)|= (1-o(1))q^n$.
    
    Note that $\mathcal{ALL}(n,P)= \cup_{a_1,a_2} \mathcal{C}_P(a_1,a_2)$, by the pigeonhole principle, there exists a choice of $a_1,a_2$ such that the size of $\mathcal{C}_P(a_1,a_2)$ is at least $\frac{|\mathcal{ALL}(n,P)|}{q(1+P/2)}$, implying that $r(\mathcal{C}_P(a_1,a_2)) \leq \log_q(1+P/2)+ \log_q q+o(1)= \log_q \log_q n+ 1-\log_q 2+ o(1)$.
\end{IEEEproof}

Therefore, we obtain the following corollary.
\begin{corollary}\label{cor:ins-bound}
  Let $\mathcal{B} \in \{\mathcal{D}_1,\mathcal{I}_1\}$, the optimal redundancy of classical $(n,2;\mathcal{B})_q$-reconstruction codes ranges from $\log_q \log_q n- \log_q 2-o(1)$ to $\log_q \log_q n+ 1-\log_q 2+o(1)$.
\end{corollary}

\section{Characterization of Two Sequences When the Hamming Distance Between Their $2$-Read vectors is Exactly $d$}\label{sec:condition}

In $\ell$-read codes, the focus is on the Hamming distance between the $\ell$-read vectors of two sequences, rather than the Hamming distance between the two sequences themselves. 
A natural question arises: what is the structure of two sequences when their $\ell$-read vectors have a Hamming distance of exactly $d$? 
Previously, Banerjee \emph{et al.} \cite{Banerjee-23-arxiv-nanopore} addressed the case of $\ell \geq 3$ and $d\leq 2$. 
In this section, we provide an answer to this question for the scenario where $\ell=2$ and for arbitrary $d$.
We first observe that the Hamming distance between any two $\ell$-read vectors is at least two for any $\ell \geq 2$.

\begin{lemma}\label{lem:HD>=2}
   Assume $\boldsymbol{x}\neq \boldsymbol{y} \in \Sigma_q^n$ and $\ell \geq 2$, we have $d_H(\mathcal{R}_{\ell}(\boldsymbol{x}),\mathcal{R}_{\ell}(\boldsymbol{y})) \geq 2$.
\end{lemma}

\begin{IEEEproof}
    If $d_H(\boldsymbol{x}, \boldsymbol{y})=1$, let $i$ be such that $x[i] \neq y[i]$, 
    it follows by the definition of $\ell$-read vectors that for any two substrings of length $\ell$ of $\boldsymbol{x}$ and $\boldsymbol{y}$ starting at the same index, their compositions are distinct if and only if they contain the $i$-th entry of $\boldsymbol{x}$ and $\boldsymbol{y}$, respectively.
    In other words, the following holds:
    \begin{equation*}
      \begin{cases}
        \mathcal{R}_{\ell}(x)[j] \neq \mathcal{R}_{\ell}(y)[j], & \mbox{if } j \in [i,i+\ell-1]; \\
        \mathcal{R}_{\ell}(x)[j]= \mathcal{R}_{\ell}(y)[j], & \mbox{otherwise}.
      \end{cases}
    \end{equation*}
    In this case, we have $d_H(\mathcal{R}_{\ell}(\boldsymbol{x}),\mathcal{R}_{\ell}(\boldsymbol{y}))= \ell \geq 2$.
    
    If $d_H(\boldsymbol{x}, \boldsymbol{y})\geq 2$, let $i, j$ be the smallest, largest indices that $\boldsymbol{x}$ and $\boldsymbol{y}$ differ, respectively, it again follows by the definition of $\ell$-read vectors that $\mathcal{R}_{\ell}(x)[i] \neq \mathcal{R}_{\ell}(y)[i]$ and $\mathcal{R}_{\ell}(x)[j+\ell-1] \neq \mathcal{R}_{\ell}(y)[j+\ell-1]$. 
    In this case, we also have $d_H(\mathcal{R}_{\ell}(\boldsymbol{x}),\mathcal{R}_{\ell}(\boldsymbol{y})) \geq 2$, thereby completing the proof.
\end{IEEEproof}

Remarkably, our proof reveals differences between the cases of $\ell \geq 3$ and $\ell=2$ (when considering two sequences with a Hamming distance of exactly one, the Hamming distance between their $\ell$-read vectors is greater than two when $\ell \geq 3$, but equals two when $\ell=2$). 
Therefore, Lemma \ref{lem:1sub} (considering the case of $\ell\geq 3$ and $d\leq 2$) does not hold for the scenario where $\ell=2$.
Below, we provide the characterization of two sequences for which the Hamming distance between their $2$-read vectors is exactly two.

\begin{theorem}\label{thm:1sub}
Assume $\boldsymbol{x} \neq \boldsymbol{y} \in \Sigma_q^n$, the Hamming distance between the $2$-read vectors of $\boldsymbol{x}$ and $\boldsymbol{y}$ is exactly two if and only if there exist two sequences $\boldsymbol{u}, \boldsymbol{v}\in \Sigma_q^{\geq 0}$, two distinct symbols $a,b \in \Sigma_q$, and an integer $t \geq 1$, such that 
\begin{equation*}
    \begin{cases}
      \boldsymbol{x}= (\boldsymbol{u}, \boldsymbol{\alpha}_t(ab), \boldsymbol{v}); \\
      \boldsymbol{y}= (\boldsymbol{u}, \boldsymbol{\alpha}_t(ba), \boldsymbol{v}).
    \end{cases}
  \end{equation*}
\end{theorem}

\begin{IEEEproof}
    Note that the `if' part can be verified directly, below we prove the `only if' part.
    Since $\boldsymbol{x} \neq \boldsymbol{y}$, there exist three sequences $\boldsymbol{u}, \boldsymbol{v}, \boldsymbol{v}' \in \Sigma_q^{\geq 0}$, two distinct symbols $a,b \in \Sigma_q$, and an integer $t \geq 1$, such that 
    \begin{equation*}
        \begin{cases}
          \boldsymbol{x}= (\boldsymbol{u}, \boldsymbol{\alpha}_t(ab), \boldsymbol{v}), \\
          \boldsymbol{y}= (\boldsymbol{u}, \boldsymbol{\alpha}_t(ba), \boldsymbol{v}'),
        \end{cases}
      \end{equation*}
      where $\{\{\alpha_t(ab)[t],v[1] \}\} \neq \{\{\alpha_t(ba)[t],v'[1] \}\}$ (in other words, at least one of $(\boldsymbol{\alpha}_t(ab),v[1])$ and $(\boldsymbol{\alpha}_t(ba),v'[1])$ is not an alternating sequence).
      To complete the proof, it remains to show that $\boldsymbol{v}= \boldsymbol{v}'$ when $\boldsymbol{v}\neq \emptyset$.
      
      Assume $|\boldsymbol{u}|=k_1$ and $|\boldsymbol{v}|=k_2>0$, it follows by the definition of $2$-read vectors that
      \begin{equation}\label{eq:d=2}
        \begin{cases}
          \mathcal{R}(\boldsymbol{x})= (\mathcal{R}(\boldsymbol{u})[1,k_1], \{\{u[k_1],a\}\}, \{\{a,b\}\}^{t-1}, \{\{\alpha_t(ab)[t],v_1 \}\}, \mathcal{R}(\boldsymbol{v})[2,k_2+1]); \\
          \mathcal{R}(\boldsymbol{y})= (\mathcal{R}(\boldsymbol{u})[1,k_1], \{\{u[k_1],b\}\}, \{\{a,b\}\}^{t-1}, \{\{\alpha_t(ba)[t],v_1' \}\}, \mathcal{R}(\boldsymbol{v}')[2,k_2+1]).
        \end{cases}
      \end{equation}
      Observe that $d_H(\mathcal{R}(\boldsymbol{x}),\mathcal{R}(\boldsymbol{y}))= d_H(\mathcal{R}(\boldsymbol{x})[1,k_1+t+1],\mathcal{R}(\boldsymbol{y})[1,k_1+t+1])+ d_H(\mathcal{R}(\boldsymbol{v})[2,k_2+1], \mathcal{R}(\boldsymbol{v}')[2,k_2+1])$ and for any $i\leq k_1+t+1$, $\mathcal{R}(x)[i]\neq \mathcal{R}(y)[i]$ holds if and only if $i\in \{k_1+1,k_1+t+1\}$, we have $\mathcal{R}(\boldsymbol{v})[2,k_2+1]= \mathcal{R}(\boldsymbol{v}')[2,k_2+1]$, i.e., $d_H(\mathcal{R}(\boldsymbol{v}), \mathcal{R}(\boldsymbol{v}')) \leq 1$, when $d_H(\mathcal{R}(\boldsymbol{x}),\mathcal{R}(\boldsymbol{y}))= 2$.
      Then by Lemma \ref{lem:1sub}, we have $d_H(\mathcal{R}(\boldsymbol{v}), \mathcal{R}(\boldsymbol{v}')) \geq 2$ when $\boldsymbol{v} \neq \boldsymbol{v}'$, thereby $\boldsymbol{v}= \boldsymbol{v}'$.
\end{IEEEproof}

Now we generalize Theorem \ref{thm:1sub} to arbitrary $d$.

\begin{theorem}\label{thm:characterization}
     Assume $\boldsymbol{x} \neq \boldsymbol{y} \in \Sigma_q^n$ and $d\geq 2$, the Hamming distance between the $2$-read vectors of $\boldsymbol{x}$ and $\boldsymbol{y}$ is exactly $d$ if and only if there exist integers $s\geq 0, t_1, \ldots,t_{s+1} \geq 1$, sequences $\boldsymbol{u}, \boldsymbol{w}, \boldsymbol{v}_1, \ldots, \boldsymbol{v}_s \in \Sigma_q^{\geq 0}$, and symbols $a_1, \ldots,a_s,b_1,\ldots,b_s \in \Sigma_q$ with $a_i \neq b_i$ for $i\in [1,s]$, such that 
     \begin{equation*}
        \begin{cases}
            \boldsymbol{x}= (\boldsymbol{u}, (\boldsymbol{\alpha}_{t_1}(a_1b_1), \boldsymbol{v}_1), \ldots, (\boldsymbol{\alpha}_{t_s}(a_sb_s), \boldsymbol{v}_s), \boldsymbol{\alpha}_{t_{s+1}}(a_{s+1}b_{s+1}), \boldsymbol{w}), \\
            \boldsymbol{y}= (\boldsymbol{u}, (\boldsymbol{\alpha}_{t_1}(b_1a_1), \boldsymbol{v}_1), \ldots, (\boldsymbol{\alpha}_{t_s}(b_sa_s), \boldsymbol{v}_s), \boldsymbol{\alpha}_{t_{s+1}}(b_{s+1}a_{s+1}), \boldsymbol{w}),
          \end{cases}
        \end{equation*}
        and the following conditions hold:
     \begin{itemize}
       \item $\{\{\alpha_{t_i}(a_ib_i)[t_i],a_{i+1}\}\} \neq \{\{\alpha_{t_i}(b_ia_i)[t_i],b_{i+1}\}\}$ (in other words, at least one of $(\boldsymbol{\alpha}_{t_i}(a_ib_i)[t_i],a_{i+1})$ and $(\alpha_{t_i}(b_ia_i)[t_i],b_{i+1})$ is not an alternating sequence) when $\boldsymbol{v}_i= \emptyset$, for any $i \in [1,s]$;
       \item $d= 2(s+1)-|\{i \in [1,s]: \boldsymbol{v}_i=\emptyset\}|$.
     \end{itemize}
\end{theorem}

\begin{IEEEproof}
    We prove the theorem by induction on $d$.
    For the base case where $d_H(\mathcal{R}(\boldsymbol{x}), \mathcal{R}(\boldsymbol{y}))= 2$, the conclusion can be inferred by Theorem \ref{thm:1sub} directly.
    Now, assuming that the conclusion is valid for $d_H(\mathcal{R}(\boldsymbol{x}), \mathcal{R}(\boldsymbol{y}))\leq d-1$, we examine the scenario where $d_H(\mathcal{R}(\boldsymbol{x}), \mathcal{R}(\boldsymbol{y}))= d \geq 3$.
    Since $\boldsymbol{x} \neq \boldsymbol{y}$, there exist three sequences $\boldsymbol{u}, \boldsymbol{x}', \boldsymbol{y}' \in \Sigma_q^{\geq 0}$, two distinct symbols $a_1,b_1 \in \Sigma_q$, and an integer $t_1 \geq 1$, such that $\{\{\alpha_{t_1}(a_1b_1)[t_1],x'[1] \}\} \neq \{\{\alpha_{t_1}(b_1a_1)[t],y'[1] \}\}$ and the following holds
    \begin{equation*}
        \begin{cases}
          \boldsymbol{x}= (\boldsymbol{u}, \boldsymbol{\alpha}_{t_1}(a_1b_1), \boldsymbol{x}'); \\
          \boldsymbol{y}= (\boldsymbol{u}, \boldsymbol{\alpha}_{t_1}(b_1a_1), \boldsymbol{y}').
        \end{cases}
      \end{equation*}
    Let $|\boldsymbol{x}'|=k$, it follows by Equation (\ref{eq:d=2}) that $d_H(\mathcal{R}(\boldsymbol{x}')[2,k+1], \mathcal{R}(\boldsymbol{y}')[2,k+1])=d-2$.
    Let $d'\triangleq d_H(\mathcal{R}(\boldsymbol{x}'), \mathcal{R}(\boldsymbol{y}'))$. Notably, $\mathcal{R}(x')[1]=\mathcal{R}(y')[1]$ holds if and only if $x'[1]=y'[1]$, implying that $d'=d-2$ when $x'[1]=y'[1]$, and $d'=d-1$ otherwise. Then by the induction hypothesis, there exist integers $s, t_2, \ldots,t_{s+1} \geq 1$, sequences $\boldsymbol{w}, \boldsymbol{v}_1, \ldots, \boldsymbol{v}_s \in \Sigma_q^{\geq 0}$, and symbols $a_2, \ldots,a_s,b_2,\ldots,b_s \in \Sigma_q$ with $a_i \neq b_i$ for $i\in [2,s]$, such that
    \begin{equation*}
      \begin{cases}
        \boldsymbol{x}'= (\boldsymbol{v}_1, (\boldsymbol{\alpha}_{t_2}(a_2b_2), \boldsymbol{v}_2), \ldots, (\boldsymbol{\alpha}_{t_s}(a_sb_s), \boldsymbol{v}_s), \boldsymbol{\alpha}_{t_{s+1}}(a_{s+1}b_{s+1}), \boldsymbol{w}), \\
        \boldsymbol{y}'= (\boldsymbol{v}_1, (\boldsymbol{\alpha}_{t_2}(b_2a_2), \boldsymbol{v}_2), \ldots, (\boldsymbol{\alpha}_{t_s}(b_sa_s), \boldsymbol{v}_s), \boldsymbol{\alpha}_{t_{s+1}}(b_{s+1}a_{s+1}), \boldsymbol{w}),
      \end{cases}
    \end{equation*}
    and the following conditions hold:
    \begin{itemize}
      \item $\{\{\alpha_{t_i}(a_ib_i)[t_i],a_{i+1}\}\} \neq \{\{\alpha_{t_i}(b_ia_i)[t_i],b_{i+1})\}\}$ when $\boldsymbol{v}_i= \emptyset$, for $i \in [2,s]$ (this condition also holds when $i=1$ since $\{\{\alpha_{t_1}(a_1b_1)[t_1],x'[1] \}\} \neq \{\{\alpha_{t_1}(b_1a_1)[t_1],y'[1] \}\}$);
      \item $d'= 2s-|\{i \in [2,s]: \boldsymbol{v}_i=\emptyset\}|$.
    \end{itemize} 
    Note that $\boldsymbol{v}_1 \neq \emptyset$ and $d'= d-2$ when $x'[1]=y'[1]$, and $\boldsymbol{v}_1= \emptyset$ and $d'= d-1$ otherwise, we can compute $d= 2(s+1)-|\{i \in [1,s]: \boldsymbol{v}_i=\emptyset\}|$. Consequently, the conclusion is valid for $d_H(\mathcal{R}(\boldsymbol{x}), \mathcal{R}(\boldsymbol{y}))= d$, thereby completing the proof.
    \end{IEEEproof}

For the case of $d\in \{3,4\}$, below we provide a more detailed description of the structure of two sequences, which will be used later.

\begin{theorem}\label{thm:d=34}
  Assume $\boldsymbol{x} \neq \boldsymbol{y} \in \Sigma_q^n$, the following two statements are true.
  \begin{itemize}
    \item The Hamming distance between the $2$-read vectors of $\boldsymbol{x}$ and $\boldsymbol{y}$ is exactly three if and only if there exist sequences $\boldsymbol{u}, \boldsymbol{v} \in \Sigma_q^{\geq 0}$, integers $t_1,t_2 \geq 1$, and symbols $a_1, b_1, a_2, b_2 \in \Sigma_q$ with $a_1\neq b_1$, $a_2\neq b_2$, such that $\{\{\alpha_{t_1}(a_1b_1)[t_1],a_2 \}\} \neq \{\{\alpha_{t_1}(b_1a_1)[t_1],b_2\}\}$ and the following holds:
        \begin{equation*}
          \begin{cases}
            \boldsymbol{x}= (\boldsymbol{u}, \boldsymbol{\alpha}_{t_1}(a_1b_1), \boldsymbol{\alpha}_{t_2}(a_2b_2), \boldsymbol{v}); \\
            \boldsymbol{y}= (\boldsymbol{u}, \boldsymbol{\alpha}_{t_1}(b_1a_1), \boldsymbol{\alpha}_{t_2}(b_2a_2), \boldsymbol{v}).
          \end{cases}
        \end{equation*}
        
    \item The Hamming distance between the $2$-read vectors of $\boldsymbol{x}$ and $\boldsymbol{y}$ is exactly four if and only if one of the following two cases occurs.
        \begin{enumerate}[(A)]
          \item There exist sequences $\boldsymbol{u}, \boldsymbol{w} \in \Sigma_q^{\geq 0}, \boldsymbol{v}  \in \Sigma_q^{\geq 1}$, integers $t_1,t_2 \geq 1$, and symbols $a_1, b_1, a_2, b_2 \in \Sigma_q$ with $a_1\neq b_1$ and $a_2\neq b_2$, such that:
        \begin{equation*}
          \begin{cases}
            \boldsymbol{x}= (\boldsymbol{u}, \boldsymbol{\alpha}_{t_1}(a_1b_1), \boldsymbol{v}, \boldsymbol{\alpha}_{t_2}(a_2b_2), \boldsymbol{w}); \\
            \boldsymbol{y}= (\boldsymbol{u}, \boldsymbol{\alpha}_{t_1}(b_1a_1), \boldsymbol{v}, \boldsymbol{\alpha}_{t_2}(b_2a_2), \boldsymbol{w}).
          \end{cases}
        \end{equation*}
          \item There exist sequences $\boldsymbol{u}, \boldsymbol{v} \in \Sigma_q^{\geq 0}$, integers $t_1,t_2,t_3 \geq 1$, and symbols $a_1,b_1, a_2,b_2, a_3,b_3 \in \Sigma_q$ with $a_1 \neq b_1$, $a_2 \neq b_2$, and $a_3 \neq b_3$, such that $\{\{\alpha_{t_1}(a_1b_1)[t_1],a_2\}\} \neq \{\{\alpha_{t_1}(b_1a_1)[t_1],b_2\}\}$, $\{\{\alpha_{t_2}(a_2b_2)[t_2],a_3\}\} \neq \{\{\alpha_{t_2}(b_2a_2)[t_2],a_3\}\}$, and the following holds:
        \begin{equation*}
          \begin{cases}
            \boldsymbol{x}= (\boldsymbol{u}, \boldsymbol{\alpha}_{t_1}(a_1b_1), \boldsymbol{\alpha}_{t_2}(a_2b_2), \boldsymbol{\alpha}_{t_3}(a_3b_3), \boldsymbol{v}); \\
            \boldsymbol{y}= (\boldsymbol{u}, \boldsymbol{\alpha}_{t_1}(b_1a_1), \boldsymbol{\alpha}_{t_2}(b_2a_2), \boldsymbol{\alpha}_{t_3}(b_3a_3), \boldsymbol{v}).
          \end{cases}
        \end{equation*}
        \end{enumerate}
    \end{itemize}
\end{theorem}

\begin{IEEEproof}
    In Theorem \ref{thm:characterization}, it is noted that $s+2 \leq d= 2(s+1)- |\{i \in [1,s]:\boldsymbol{v}_i=\emptyset\}| \leq 2(s+1)$.
    \begin{itemize}
      \item When $s=0$, we have $d=2$.
      \item When $s=1$, we have $d\in \{3,4\}$. Moreover, $d=3$ holds if and only if $\boldsymbol{v}_1= \emptyset$, and $d=4$ holds if and only if $\boldsymbol{v}_1\neq \emptyset$.
      \item When $s=2$, we have $d\geq 4$. Moreover, $d=4$ holds if and only if $\boldsymbol{v}_1= \boldsymbol{v}_2= \emptyset$.
      \item When $s \geq 3$, we have $d \geq 5$.
    \end{itemize}
    Again by Theorem \ref{thm:characterization}, the conclusion follows.
\end{IEEEproof}

By knowing when $2$-read vectors have a Hamming distance of exactly $d$, we can investigate the bounds and constructions of $2$-read $(n,d)_q$-codes.
In the subsequent three sections, we consider the scenario where $d \in \{3,4,5\}$.
Moreover, when $d=3$, the focus will be on any read length rather than just $\ell=2$, since previous work \cite{Banerjee-23-arxiv-nanopore} has determined the condition for $\ell$-read vectors with $\ell\geq 3$ to have a Hamming distance of exactly two.

\section{$\ell$-read codes with a minimum distance of three}\label{sec:d=3}

In \cite{Banerjee-23-arxiv-nanopore}, Banerjee \emph{et al.} focused on $\ell$-read $(n,3)_q$-codes with the constraint $\ell \geq 3$ and showed that the optimal redundancy of the codes ranges from $\log_q\log_q n- \log_q 2-o(1)$ to $\log_q \log_q n+ \log_q 2(q-1)+1+o(1)$. 
In this section, we first improve their upper bound by constructing such a code with $\log_q\log_q n-\log_q 2+o(1)$ bits of redundancy, where the redundancy is asymptotically optimal as the difference between it and the theoretical lower bound becomes negligible as the code length increases.
We then study $2$-read $(n,3)_q$-codes and demonstrate that the optimal redundancy of the codes ranges from $\log_q\log_q n-o(1)$ to  $\log_q\log_q n+ 1-\log_q2+o(1)$. 
Furthermore, we establish an equivalence between $2$-read $(n,3)_q$-codes and classical $(n,2;\mathcal{I}_1)_q$-reconstruction codes and improve the lower bound on the redundancy of classical $(n,2;\mathcal{I}_1)_q$-reconstruction codes as well as the upper bound on the redundancy of classical $(n,2;\mathcal{D}_1)_q$-reconstruction codes.

\subsection{Upper Bound on $\ell$-Read Codes with a Minimum Distance of Three where $\ell \geq 3$}

In this subsection, assume $\ell\geq 3$, we present a construction of $\ell$-read $(n,3)_q$-codes asymptotically achieving the lower bound on the redundancy derived in \cite{Banerjee-23-arxiv-nanopore}, implying that our construction is asymptotically optimal.
To do so, we first introduce the notion of a `good' sequence, which will be important for the code construction.

\begin{definition}
  Assume $\ell\geq 3$, $\boldsymbol{x} \in \Sigma_q^n$ is referred to as a \emph{good} sequence if $((x[i],x[i+1]),(x[i+\ell],x[i+\ell+1]), \ldots,(x[i+t\ell]x[i+t\ell+1]))$ is not an alternating sequence for any $i \in [1,n-t\ell-1]$, when $t \geq \frac{\log_q n+ \log_q\log_q n}{2}-1$. 
\end{definition}

The following lemma shows that as $n \rightarrow \infty$, almost all sequences (a fraction $1-o(1)$) are good.
\begin{lemma}
  The number of good sequences in $\Sigma_q^n$ is $(1-o(1))q^n$ as $n \rightarrow \infty$.
\end{lemma}

\begin{IEEEproof}
    Choose a sequence $\boldsymbol{x}\in \Sigma_q^n$ uniformly at random, the probability that $\boldsymbol{x}$ is good can be calculated as:
    \begin{align*}
    \mathbbm{P} (\boldsymbol{x} \text{ is good})
    & \geq 1- n \frac{q(q-1)}{q^{\log_q n+\log_q\log_q n}}
    \geq 1- \frac{q(q-1)}{\log_q n}= 1-o(1).
    \end{align*}
    Then the conclusion follows.
\end{IEEEproof}

Based on this, we propose a code construction in the following theorem that only includes good sequences satisfying an additional inversion number condition.

\begin{theorem}\label{thm:code}
  Assume $\ell\geq 3$. Let $P=\lceil \frac{\log_q n + \log_q \log_q n}{2} \rceil$ and $a \in [0,P-1]$, the code
  \begin{equation*}
    \mathcal{C}_{3,3}^{\ell,d} \triangleq 
    \{\boldsymbol{x} \in \Sigma_q^n: \boldsymbol{x} \text{ is good},
        ~\mathrm{Inv}(\boldsymbol{x})\equiv a \pmod{P}\}
  \end{equation*}
   is an $\ell$-read $(n,3)_q$-code.
   Moreover, by the pigeonhole principle, there exists $a \in [0,P-1]$ such that $r(\mathcal{C}_{3,3}^{\ell,d}) \leq \log_q\log_q P+o(1)= \log_q\log_q n-\log_q 2+o(1)$.
\end{theorem}

\begin{IEEEproof}
    Assume $\mathcal{C}_{3,3}^{\ell,d}$ is not an $\ell$-read $(n,3)_q$-code, then there exist two distinct sequences
    $\boldsymbol{x}$ and $\boldsymbol{y}$ in $\mathcal{C}_{3,3}^{\ell,d}$ such that $d_H(\mathcal{R}_{\ell}(\boldsymbol{x}),\mathcal{R}_{\ell}(\boldsymbol{y})) \leq 2$. 
    It then follows by Lemma \ref{lem:1sub} that 
  \begin{equation*}
    \begin{cases}
      \boldsymbol{x}= (\boldsymbol{u}, (a, b, \boldsymbol{v}_1), (a,b, \boldsymbol{v}_2), \ldots, (a,b, \boldsymbol{v}_{t}),a,b, \boldsymbol{w}); \\
      \boldsymbol{y}= (\boldsymbol{u}, (b, a, \boldsymbol{v}_1), (b,a, \boldsymbol{v}_2),\ldots, (b,a, \boldsymbol{v}_{t}),b,a, \boldsymbol{w}),
    \end{cases}
  \end{equation*} 
  for some $\boldsymbol{u}, \boldsymbol{w} \in \Sigma_q^{\geq 0}, \boldsymbol{v}_1, \ldots, \boldsymbol{v}_t \in \Sigma_q^{\ell-2}$, where $t\geq 0$, and $a\neq b \in \Sigma_q$ (without loss of generality assume $a>b$).
  
  Let $\boldsymbol{x}_0=\boldsymbol{x}$, $\boldsymbol{x}_{t+1}=\boldsymbol{y}$, and $\boldsymbol{x}_i\triangleq (\boldsymbol{u}, (b,a, \boldsymbol{v}_1), \ldots, (b,a, \boldsymbol{v}_{i}), (a,b, \boldsymbol{v}_{i+1}), \ldots, (a,b, \boldsymbol{v}_{t}), a,b, \boldsymbol{w})$ for any $i\in [1,t]$, we may compute $\mathrm{Inv}(\boldsymbol{x}_i)- \mathrm{Inv}(\boldsymbol{x}_{i+1})=1$ for any $i \in [0,t]$.
  Therefore, we get
  \begin{align*}
    \mathrm{Inv}(\boldsymbol{x})- \mathrm{Inv}(\boldsymbol{y})
    = \sum_{i=0}^t (\mathrm{Inv}(\boldsymbol{x}_i)- \mathrm{Inv}(\boldsymbol{x}_{i+1}))
    = t+1.
  \end{align*}
  Since $\boldsymbol{x}$ is good, we have $t< \frac{\log_q n+\log_q\log_q n}{2}-1\leq P-1$, implying that $0<t+1<P$.
  This contradicts the condition that $\mathrm{Inv}(\boldsymbol{x})\equiv \mathrm{Inv}(\boldsymbol{y}) \pmod{P}$.
  Therefore, $\mathcal{C}_{3,3}^{\ell,d}$ is an $\ell$-read $(n,3)_q$-code.
  Furthermore, the redundancy of $\mathcal{C}_{3,3}^{\ell,d}$ can be determined by applying the pigeonhole principle directly, thereby completing the proof.
\end{IEEEproof}

Combining this with Lemma \ref{lem:l>=3}, the following holds.
\begin{corollary}\label{cor:d=3,l>=3}
  For $\ell\geq 3$, the optimal redundancy of $\ell$-read $(n,3)_q$-codes asymptotically approaches to $\log_q\log_q n-\log_q 2$.
\end{corollary}


\subsection{Bounds on $2$-Read Codes with a Minimum Distance of Three}

For any two distinct sequences $\boldsymbol{x}$ and $\boldsymbol{y}$, it can be easily checked by Lemma \ref{lem:indel} and Theorem \ref{thm:1sub} that $d_H(\mathcal{R}(\boldsymbol{x}),\mathcal{R}(\boldsymbol{y}))= 2$ holds if and only if $|\mathcal{I}_1(\boldsymbol{x}) \cap \mathcal{I}_1(\boldsymbol{y})|=2$.
Therefore, a code $\mathcal{C} \subseteq \Sigma_q^n$ is $2$-read $(n,3)_q$-code if and only if it is a classical $(n,2;\mathcal{I}_1)_q$-reconstruction code.
Then by Corollary \ref{cor:ins-bound}, the optimal redundancy of $2$-read $(n,3)_q$-codes ranges from $\log_q \log_q n- \log_q 2- o(1)$ to $\log_q \log_q n+ 1-\log_q 2+ o(1)$.
In this subsection, we improve the lower bound to $\log_q\log_q n-o(1)$ 
using an additional piece of information compared to the method in \cite{Chrisnata-22-IT-reconstr-del}.
Specifically, for any two sequences $\boldsymbol{x}$ and $\boldsymbol{y}$, we utilize the fact that when the Hamming distance between them is one, the Hamming distance between their $2$-read vectors is two, or equivalently the size of the intersection between their single-insertion balls is two.

Considering the graph $\mathcal{G}= (V,E)$ where $V= \Sigma_q^n$ and $E=\{\{\boldsymbol{x}, \boldsymbol{y}\}\in V^2: d_H(\mathcal{R}(\boldsymbol{x}),\mathcal{R}(\boldsymbol{y}))= 2\}$, an \emph{independent set} of $\mathcal{G}$ is a subset of $V$ in which no pair of sequences belongs to $E$ and the \emph{independence number} of $\mathcal{G}$ is the cardinality of the largest independent set of $\mathcal{G}$, denoted as $\alpha(\mathcal{G})$.
Therefore, any independent set of $\mathcal{G}$ is an $\ell$-read $(n,3)_q$-code, and the independence number of $\mathcal{G}$ is an upper bound on the size of $\ell$-read $(n,3)_q$-codes.
Consequently, the optimal redundancy of $\ell$-read $(n,3)_q$-codes is lower bounded by $n - \log_q \alpha(\mathcal{G})$.

\begin{definition}
$\mathcal{Q}$ is referred to as a \emph{clique cover} of $\mathcal{G}$, if it is a collection of cliques such that each vertex in $\mathcal{G}$ belongs to some clique in $\mathcal{Q}$.
\end{definition}

\begin{lemma}[\cite{Knuth-94-EJC-clique}]\label{lem:clique-cover}
If $\mathcal{Q}$ is a clique cover of $\mathcal{G}$, then the number of cliques in $\mathcal{Q}$ is no less than the independence number of $\mathcal{G}$, i.e., $|\mathcal{Q}| \geq \alpha(\mathcal{G})$.
\end{lemma}

Let $\mathcal{Q}$ be a clique cover of $\mathcal{G}$, it follows by Lemma \ref{lem:clique-cover} that the optimal redundancy of $2$-read $(n,3)_q$-codes is lower bounded by $n- \log_q |\mathcal{Q}|$.
In the following, we present a clique cover of $\mathcal{G}$, and thus give a lower bound on the redundancy of $2$-read $(n,3)_q$-codes.

Assume $t>0$ is an integer, let $\boldsymbol{g}_{i,t} \triangleq (\boldsymbol{\alpha}_{2t}(01)_{[1,i]}, \boldsymbol{\alpha}_{2t}(10)_{[i+1,2t]})$ for any $i \in [0,2t]$.
We observe that the Hamming distance between $\mathcal{R}(\boldsymbol{g}_{i,t})$ and $\mathcal{R}(\boldsymbol{g}_{j,t})$ is two for any $0\leq i\neq j\leq 2t$.

\begin{lemma}\label{lem:g_i}
   Assume $t>0$ is an integer, we have $d_H(\mathcal{R}(\boldsymbol{g}_{i,t}), \mathcal{R}(\boldsymbol{g}_{j,t}))=2$ for any $0\leq i\neq j\leq 2t$.
\end{lemma}

\begin{IEEEproof}
    Without loss of generality assume $i<j$, then we have
    \begin{equation*}
        \begin{cases}
          \boldsymbol{g}_{i,t}= (\boldsymbol{\alpha}_{2t}(01)_{[1,i]}, \boldsymbol{\alpha}_{2t}(10)_{[i+1,2t]})= (\boldsymbol{\alpha}_{2t}(01)_{[1,i]}, \boldsymbol{\alpha}_{2t}(10)_{[i+1,j]}, \boldsymbol{\alpha}_{2t}(10)_{[j+1,2t]});\\
          \boldsymbol{g}_{j,t}= (\boldsymbol{\alpha}_{2t}(01)_{[1,j]}, \boldsymbol{\alpha}_{2t}(10)_{[j+1,2t]})= (\boldsymbol{\alpha}_{2t}(01)_{[1,i]}, \boldsymbol{\alpha}_{2t}(01)_{[i+1,j]}, \boldsymbol{\alpha}_{2t}(10)_{[j+1,2t]}).
        \end{cases}
    \end{equation*}
    It follows by Theorem \ref{thm:1sub} that $d_H(\mathcal{R}(\boldsymbol{g}_{i,t}), \mathcal{R}(\boldsymbol{g}_{j,t}))=2$.
\end{IEEEproof}

With the help of $\boldsymbol{g}_{i,t}$, we present the following construction.

\begin{construction}
Assume $t\geq 1$ and $n=2m t+n'$, where $n'<2t$. 
Let $\Lambda= \{\boldsymbol{g}_{i,t}: i \in [0,2t] \}$ and $\tilde{\Lambda}= \Sigma_q^{2t} \setminus \Lambda$.
Now we define two types of sets, which will be proven to be cliques later.
\begin{itemize}
  \item The first type of sets are singletons $S_{\boldsymbol{x}} = \{\boldsymbol{x}\}$ for $\boldsymbol{x} \in \tilde{\Lambda}^{m} \times \Sigma_q^{n'}$;
  \item Let $\Gamma= \{(k,\boldsymbol{u}, \boldsymbol{w}): k \in [1,m], \boldsymbol{u} \in \tilde{\Lambda}^{k-1}, \boldsymbol{w} \in \Sigma_q^{n-2tk} \}$. 
      The second type of sets is defined as $\mathcal{Q}_{\boldsymbol{z}}= \{(\boldsymbol{u}, \boldsymbol{g}_{i,t}, \boldsymbol{w}): i \in [0,2t] \}$ for $\boldsymbol{z}= (k,\boldsymbol{u}, \boldsymbol{w}) \in \Gamma$.
\end{itemize}
Finally, let $\mathcal{Q}(t)\triangleq \{S_{\boldsymbol{x}}: \boldsymbol{x} \in \tilde{\Lambda}^m \times \Sigma_q^{n'}\} \cup \{Q_{\boldsymbol{z}}: \boldsymbol{z} \in \Gamma\}$. 
\end{construction}

In the subsequent lemmas, we will demonstrate that $\mathcal{Q}(t)$ is a clique cover of $\mathcal{G}$ and determine the size of $\mathcal{Q}(t)$ as well.

\begin{lemma}
$\mathcal{Q}(t)$ is a clique cover of $\mathcal{G}$.
\end{lemma}

\begin{IEEEproof}
We first prove that $\mathcal{Q}(t)$ is a set of cliques.
Since $S_{\boldsymbol{x}}$ only contains one element, $S_{\boldsymbol{x}}$ is a clique for $\boldsymbol{x} \in \tilde{\Lambda}^m \times \Sigma_q^{n'}$. 
By Lemma \ref{lem:g_i}, it can be easily verified that $\mathcal{Q}_{\boldsymbol{z}}$ is also a clique for $\boldsymbol{z} \in \Gamma$.
Therefore, all elements in $\mathcal{Q}(t)$ are cliques.

To complete the proof, it suffices to show that each $\boldsymbol{x} \in \Sigma_q^n$ belongs to some clique in $\mathcal{Q}(t)$. 
\begin{itemize}
  \item If $\boldsymbol{x} \in \tilde{\Lambda}^m \times \Sigma_q^{n'}$, we have $\boldsymbol{x} \in \mathcal{Q}(t)$.
  \item If $\boldsymbol{x} \not\in \tilde{\Lambda}^m \times \Sigma_q^{n'}$, let $\boldsymbol{x}_i \triangleq \boldsymbol{x}[2(i-1) t+1,2i t]$ for $i \in [1,m]$. Assume $k$ is the smallest index that $\boldsymbol{x}_k \not\in \tilde{\Lambda}$, then $\boldsymbol{x}[1,2(k-1)t] \in \tilde{\Lambda}^{k-1}$.
      Let $\boldsymbol{z}= (k, \boldsymbol{x}[1,2(k-1)t],\boldsymbol{x}[2kt+1,n])$, then $\boldsymbol{z} \in \Gamma$ and $\boldsymbol{x} \in \mathcal{Q}_{\boldsymbol{z}}$, thereby completing the proof.
\end{itemize}
\end{IEEEproof}

\begin{lemma}
The number of cliques in $\mathcal{Q}(t)$ is $\frac{q^n}{2t+1} ( 1+2t( 1-\frac{2t+1}{q^{2t}} )^{m} )$.
\end{lemma}

\begin{IEEEproof}
The number of the first type of cliques is $( q^{2t}-(2t+1))^{m} q^{n-2tm}$, and that of the second type of cliques is $\sum_{k=1}^m (q^{2t}- (2t+1) )^{k-1} q^{n-2tk}$. Therefore, we may compute

\begin{equation*}
    \begin{aligned}
        |\mathcal{Q}(t)|
            &= ( q^{2t}- (2t+1) )^{m}q^{n-2tm} +  \sum_{k=1}^m ( q^{2t}- (2t+1) )^{k-1} q^{n-2tk} \\
            &= q^n (1- \frac{2t+1}{q^{2t}} )^{m} + q^{n-2t} \sum_{k=1}^m
            ( 1-\frac{2t+1}{q^{2t}} )^{k-1}\\
            &= q^n (1- \frac{2t+1}{q^{2t}} )^{m} + \frac{q^n}{2t+1} ( 1 - ( 1-\frac{2t+1}{q^{2t}} )^m ) \\
            &= \frac{q^n}{2t+1} ( 1+2t( 1-\frac{2t+1}{q^{2t}} )^{m} ).
    \end{aligned}
\end{equation*}
\end{IEEEproof}

Since our objective is to construct a clique cover of $\mathcal{G}$ with the largest size, we need to select a suitable value of $t$ to optimize the size of $Q(t)$.

\begin{lemma}\label{lem:size}
  Set $t= \lfloor \frac{1}{2}(\log_q n- \log_q (2 \ln (\log_q n))) \rfloor$, we have $|\mathcal{Q}(t)|= (1+o(1))\frac{q^n}{\log_q n}$ as $n \rightarrow \infty$.
\end{lemma}

\begin{IEEEproof}
    We may compute $q^{2t} \leq \frac{n}{2 \ln (\log_q n)}$.
    It follows by $1-x \leq e^{-x}$ that
    \begin{align*}
      t ( 1-\frac{2t+1}{q^{2t}})^{\lfloor \frac{n}{t} \rfloor}
        &\leq t \exp{ \left\{-\frac{2t+1}{q^{2t}}(\frac{n}{t}-1) \right\}} \\
        &\leq t \exp{ \left\{-\frac{n}{q^{2t}} \right\}} \\
        &\leq t \exp{ \left\{-2 \ln (\log_q n) \right\}} \rightarrow 0.
    \end{align*}
    Then the conclusion follows.
\end{IEEEproof}

Since the optimal redundancy of $2$-read $(n,3)_q$-codes is lower bounded by $n -\log_q |\mathcal{Q}(t)|$, by setting $t= \lfloor \frac{1}{2}(\log_q n- \log_q (2 \ln (\log_q n))) \rfloor$, the following theorem holds.

\begin{theorem}\label{thm:new-bound}
Let $\mathcal{C}$ be a $2$-read $(n,3)_q$-code or a classical $(n,2;\mathcal{I}_1)_q$-reconstruction code, we have $r(\mathcal{C}) \geq \log_q \log_q n- o(1)$.
\end{theorem}

Combining this with Lemma \ref{lem:ins_bound}, the following holds.
\begin{corollary}\label{cor:d=3,l=2}
  The optimal redundancy of $2$-read $(n,3)_q$-codes and classical $(n,2;\mathcal{I}_1)_q$-reconstruction codes ranges from $\log_q\log_q n-o(1)$ to $\log_q\log_q n+1-\log_q 2+o(1)$.
\end{corollary}

Remarkably, when $q=2$, the code constructed in Lemma \ref{lem:all} is a $2$-read $(n,3)_q$-code, or equivalently a classical $(n,2;\mathcal{I}_1)_q$-reconstruction code, with asymptotically optimal redundancy.

\subsection{Upper Bound on Classical $(n,2;\mathcal{D}_1)_q$-Reconstruction Codes}

It is well known that an equivalence exists between classical single-insertion correcting codes and classical single-deletion correcting codes \cite{Levenshtein-66-SPD-1D}. 
However, this equivalence may not hold in the reconstruction model, especially when using two noisy reads.
In the previous subsection, we improve the lower bound on the redundancy of classical $(n,2;\mathcal{I}_1)_q$-reconstruction codes.
In this subsection, we improve the upper bound on the redundancy of classical $(n,2;\mathcal{D}_1)_q$-reconstruction codes by constructing such a code.

The following definition is important for our code construction.
\begin{definition}
    For any $\boldsymbol{x} \in \Sigma_q^n$, let $\mathcal{O}(\boldsymbol{x})= (x[1],x[3],\ldots,x[2\lceil n/2 \rceil -1]) \in \Sigma_q^{\lceil n/2 \rceil}$ be its \emph{odd sequence}, and $\mathcal{E}(\boldsymbol{x})= (x[2],x[4],\ldots,x[2\lfloor n/2 \rfloor]) \in \Sigma_q^{\lfloor n/2 \rfloor}$ be its \emph{even sequence}.
\end{definition}

\begin{theorem}\label{thm:del-recons}
Assume $P=\log_q n+\log_q \log_q n$ and $p$ is the smallest prime larger than $\lceil P/2 \rceil$.
Let $m= \min\{p, (q-1)\lceil P/2 \rceil+1\}$, then for any $a \in [0,m-1]$, the code
\begin{equation*}
  \mathcal{C}_{del}= 
  \{\boldsymbol{x} \in \mathcal{ALL}(n,P): \mathrm{VT}^{(0)}(\mathcal{O}(\boldsymbol{x})) \equiv a \pmod{m} \}
\end{equation*}
is a classical $(n,2;\mathcal{D}_1)_q$-reconstruction code.
Moreover, by the pigeonhole principle, there exists $a \in [0,m]$ such that $r(\mathcal{C}_{del}) \leq \log_q\log_q m +o(1)= \log_q\log_q n+ \min\{\log_q(q-1)-\log_q 2,0\}+o(1)$.
\end{theorem}

\begin{IEEEproof}
    Assume $\mathcal{C}_{del}$ is not a classical $(n,2;\mathcal{D}_1)_q$-reconstruction code, then there exist two distinct sequences $\boldsymbol{x}$ and $\boldsymbol{y}$ in $\mathcal{C}_{del}$ such that $|\mathcal{D}_1(\boldsymbol{x}) \cap \mathcal{D}_1(\boldsymbol{y})| =2$.
    It follows by Lemma \ref{lem:indel} that
    \begin{equation*}
        \begin{cases}
          \boldsymbol{x}= (\boldsymbol{u}, \boldsymbol{\alpha}_t(ab), \boldsymbol{v}), \\
          \boldsymbol{y}= (\boldsymbol{u}, \boldsymbol{\alpha}_t(ba), \boldsymbol{v}),
        \end{cases}
    \end{equation*}
    for some $\boldsymbol{u}, \boldsymbol{v} \in \Sigma_q^{\geq 0}$, $a\neq b \in \Sigma_q$, and $t \geq 2$.
     Assume $|\boldsymbol{u}|=u$, we consider the odd sequences of $\boldsymbol{x}$ and $\boldsymbol{y}$ and distinguish between the following four cases based on the parity of $u$ and $t$.
     \begin{itemize}
       \item If $2|u$ and $2|t$, we have
            \begin{equation*}
                \begin{cases}
                  \mathcal{O}(\boldsymbol{x})= (\mathcal{O}(\boldsymbol{u}), a^{t/2}, \mathcal{O}(\boldsymbol{v})); \\
                  \mathcal{O}(\boldsymbol{y})= (\mathcal{O}(\boldsymbol{u}), b^{t/2}, \mathcal{O}(\boldsymbol{v})).
                \end{cases}
            \end{equation*}
       \item If $2 \nmid u$ and $2|t$, we have
            \begin{equation*}
                \begin{cases}
                  \mathcal{O}(\boldsymbol{x})= (\mathcal{O}(\boldsymbol{u}), b^{t/2}, \mathcal{E}(\boldsymbol{v})); \\
                  \mathcal{O}(\boldsymbol{y})= (\mathcal{O}(\boldsymbol{u}), a^{t/2}, \mathcal{E}(\boldsymbol{v})).
                \end{cases}
            \end{equation*}
       \item If $2|u$ and $2 \nmid t$, we have
            \begin{equation*}
                \begin{cases}
                  \mathcal{O}(\boldsymbol{x})= (\mathcal{O}(\boldsymbol{u}), a^{\lceil t/2 \rceil}, \mathcal{E}(\boldsymbol{v})); \\
                  \mathcal{O}(\boldsymbol{y})= (\mathcal{O}(\boldsymbol{u}), b^{\lceil t/2 \rceil}, \mathcal{E}(\boldsymbol{v})).
                \end{cases}
            \end{equation*}
       \item If $2 \nmid u$ and $2 \nmid t$, we have
            \begin{equation*}
                \begin{cases}
                  \mathcal{O}(\boldsymbol{x})= (\mathcal{O}(\boldsymbol{u}), b^{\lfloor t/2 \rfloor}, \mathcal{O}(\boldsymbol{v})); \\
                  \mathcal{O}(\boldsymbol{y})= (\mathcal{O}(\boldsymbol{u}), a^{\lfloor t/2 \rfloor}, \mathcal{O}(\boldsymbol{v})).
                \end{cases}
            \end{equation*}
     \end{itemize}
     Since $\boldsymbol{x} \in \mathcal{ALL}(n,P)$, we have $2 \leq t \leq P$. 
     Then we may compute
     \begin{align*}
       |\mathrm{VT}^{(0)}(\mathcal{O}(\boldsymbol{x}))- \mathrm{VT}^{(0)}(\mathcal{O}(\boldsymbol{y}))| \in \{|b-a|\lceil t/2 \rceil, |b-a|\lfloor t/2\rfloor\}.
     \end{align*}
     This contradicts the condition that $\mathrm{VT}^{(0)}(\mathcal{O}(\boldsymbol{x})) \equiv \mathrm{VT}^{(0)}(\mathcal{O}(\boldsymbol{y})) \pmod{m}$.
     Thus, $\mathcal{C}_{del}$ is a classical $(n,2;\mathcal{D}_1)_q$-reconstruction code.
     Furthermore, since $p$ is the smallest prime larger than $\lceil P/2 \rceil$, we have $(\frac{1}{2}+o(1))\log_q n \leq p \leq (1+o(1))\log_q n$.
     Then the redundancy of $\mathcal{C}_{del}$ can be determined by applying the pigeonhole principle directly, thereby completing the proof.
\end{IEEEproof}

Combining this with Lemma \ref{lem:ins_bound}, the following holds.
\begin{corollary}\label{cor:del-recon}
  The optimal redundancy of classical $(n,2;\mathcal{D}_1)_q$-reconstruction codes ranges from $\log_q\log_q n-\log_q 2 -o(1)$ to $\log_q\log_q n+\min\{\log_q (q-1)-\log_q 2,0\}+o(1)$.
\end{corollary}

Remarkably, our code constructed in Theorem \ref{thm:del-recons} is better than that in Lemma \ref{lem:all}, which has redundancy $\log_q\log_q n+ 1-\log_q 2+o(1)$, and is asymptotically optimal when $q=2$.
Furthermore, it follows by Corollaries \ref{cor:d=3,l=2} and \ref{cor:del-recon} that when $q=2$, the optimal redundancy of classical $(n,2;\mathcal{D}_1)_q$-reconstruction codes is less than that of classical $(n,2;\mathcal{I}_1)_q$-reconstruction codes.


\section{$2$-read codes with a minimum distance of four}\label{sec:d=4}

Since any $2$-read $(n,4)_q$-code is also a $2$-read $(n,3)_q$-code, it follows by the previous section that the optimal redundancy of $2$-read $(n,4)_q$-codes is lower bounded by $\log_q\log_q n-o(1)$.
In this section, we focus on the construction of $2$-read $(n,4)_q$-codes.
To do so, we first introduce the following indicator sequence, which plays a crucial role in our code construction.

\begin{definition}
For any $\boldsymbol{x} \in \Sigma_q^n$, we define its \emph{indicator sequence} as $\mathbbm{1}(\boldsymbol{x})\in \Sigma_q^n$, where $\mathbbm{1}(x)[i]\triangleq x[i]+x[i-1] \pmod{q}$ for $i \in [1,n]$. 
Note that $x[1]= \mathbbm{1}(x)[1]$ and $x[i]= \mathbbm{1}(x)[i]-x[i-1] \pmod{q}$ for $i \in [2,n]$, indicating that there is a one to one correspondence between $\boldsymbol{x}$ and $\mathbbm{1}(\boldsymbol{x})$.

\end{definition}

An important observation is that for two sequences, the distance between their indicator sequences is upper bounded by the distance between their $2$-read vectors, as shown below.

\begin{lemma}\label{lem:HD}
Assume $\boldsymbol{x} \neq \boldsymbol{y} \in \Sigma_q^n$ with $d_H(\mathcal{R}(\boldsymbol{x}),\mathcal{R}(\boldsymbol{y}))= d \geq 2$, then $1 \leq d_H(\mathbbm{1}(\boldsymbol{x}),\mathbbm{1}(\boldsymbol{y})) \leq d$.
\end{lemma}

\begin{IEEEproof}
Since $x[i-1]+x[i]=y[i-1]+y[i]$ when $\{\{x[i-1],x[i]\}\}= \{\{y[i-1],y[i]\}\}$, we have $d_H(\mathbbm{1}(\boldsymbol{x}),\mathbbm{1}(\boldsymbol{y})) \leq d_H(\mathcal{R}(\boldsymbol{x}),\mathcal{R}(\boldsymbol{y}))$.
Moreover, it follows by $\boldsymbol{x}\neq \boldsymbol{y}$ that $d_H(\mathbbm{1}(\boldsymbol{x}),\mathbbm{1}(\boldsymbol{y})) \geq 1$, thereby completing the proof.
\end{IEEEproof}

Lemma \ref{lem:HD} allows us to construct $2$-read $(n,4)_q$-codes by using classical $(n,4)_q$-codes.
However, this would demand $O(\log_q n)$ bits of redundancy and ignore an important positional information, which we will now explain.
Let $P>0$ be an integer and $\boldsymbol{x}\neq \boldsymbol{y} \in \mathcal{ALL}(n,\frac{P-1}{2})$ with $d_H(\mathcal{R}(\boldsymbol{x}),\mathcal{R}(\boldsymbol{y}))\leq 3$, it can be easily verified, based on Theorems \ref{thm:1sub} and \ref{thm:d=34}, that the distance between the largest and smallest indices that $\mathbbm{1}(\boldsymbol{x})$ and $\mathbbm{1}(\boldsymbol{y})$ differ is at most $\max\{t,t_1+t_2\} \leq P-1$.
Taking advantage of this additional positional information, we can construct $2$-read $(n,4)_q$-codes with less redundancy compared to using classical $(n,4)_q$-codes directly.

\subsection{Construction of Classical $(n,d)_q$-Codes with Additional Positional Information}

In this subsection, we explore a general model in which there are no pairs of sequences such that the Hamming distance between them is less than $d$ when the distance between the largest and smallest indices that their indicator sequences differ is less than $P$.

\begin{theorem}\label{thm:bounded}
Let $p$ be the smallest prime satisfying $p \geq \max\{P,q\}$. 
Let $a_{0} \in [0,p_0-1]$ and $a_1,a_2,\ldots,a_{d-2} \in[0, p_i-1]$ where $p_0= (d-1)(q-1)+1$ and $p_i=p$ for $i \in [1,d-2]$. 
Define the code
\begin{align*}
  \mathcal{C}_{q,d,P}= 
  & \{\boldsymbol{x}\in  \Sigma_q^{n}: \mathrm{VT}^{(i)}(\boldsymbol{x}) \equiv a_i \pmod{p_i} \text { for } i \in [0, d-2] \}.
\end{align*}
For any two distinct sequences $\boldsymbol{x},\boldsymbol{y} \in \mathcal{C}_{q,d,P}$, if the distance between the smallest and largest indices that they differ is less than $P$, the Hamming distance between them is at least $d$.
Moreover, by the pigeonhole principle, there exists a choice of parameters, such that $r(\mathcal{C}_{q,d,P}) \leq (d-2)\log_q (2\max\{P,q\})+ \log_q ((d-1)(q-1)+1)$.
\end{theorem}

\begin{IEEEproof}
    Assume $d_H(\boldsymbol{x},\boldsymbol{y})<d$, then there exist $d-1$ indices $s+t_1,s+t_2,\ldots,s+t_{d-1} \in [1,n]$ with $1= t_1<t_2<\cdots<t_{d-1} \leq P$ such that
    $x[i]= y[i]$ for $i \not\in \{s+t_1,s+t_2,\ldots,s+t_{d-1}\}$, we will show that $\boldsymbol{x}=\boldsymbol{y}$.
    Once this is established, the theorem holds.
    
    Let $b_i \triangleq x[s+t_i]-y[s+t_i] \in [1-q,q-1]$ for $i \in [1,d-1]$.
    Firstly, it follows by $\mathrm{VT}^{(0)}(\boldsymbol{x})- \mathrm{VT}^{(0)}(\boldsymbol{y})= \sum_{i=1}^{d-1} b_i \equiv 0 \pmod{p_0}$ that $\sum_{i=1}^{d-1} b_i= 0$.
    Then for any $k \in [1,d-2]$, if $\sum_{i=1}^{d-1}b_i t_i^{k'} \equiv 0 \pmod{p}$ for $k' \in [0,k-1]$, we may compute
    \begin{equation*}
    \begin{aligned}
      \mathrm{VT}^{(k)}(\boldsymbol{x})-\mathrm{VT}^{(k)}(\boldsymbol{y})
      &= \sum_{i=1}^{d-1} b_i (s+t_i)^k \\
      &= \sum_{i=1}^{d-1} b_i \sum_{j=0}^k \binom{k}{j} s^j t_i^{k-j} \\
      &= \sum_{j=0}^k \binom{k}{j} s^j \sum_{i=1}^{d-1} b_i t_i^{k-j} \\
      &\equiv \sum_{i=1}^{d-1} b_i t_i^k \pmod{p}.
    \end{aligned}
    \end{equation*}
    This implies that $\sum_{i=1}^{d-1} b_i t_i^k\equiv 0 \pmod{p}$.
    Therefore, we get
    \begin{equation}\label{eq:VTk}
      \sum_{i=1}^{d-1} b_i t_i^k \equiv 0 \pmod{p} \text{ for } k \in [0,d-2].
    \end{equation}
    In other words, $(b_1,b_2,\ldots,b_{d-1})$ is a solution for $\boldsymbol{z} A= 0^{d-1} \pmod{p}$, where
    \begin{align*}
    A= \begin{pmatrix}
        1 & t_{1} & \cdots & t_{1}^{d-2} \\
        1 & t_{2} & \cdots & t_{2}^{d-2} \\
        \vdots & \vdots & \ddots & \vdots \\
        1 & t_{d-1} & \cdots & t_{d-1}^{d-2} 
       \end{pmatrix}.
    \end{align*}
    Since $A$ is invertible, we get $b_i\equiv 0 \pmod{p}$ for $i \in [1,d-1]$.
    It then follows by $p \geq q$ and $b_i \in [1-q,q-1]$ that $b_i=0$, i.e., $x[s+t_i]=y[s+t_i]$, for $i \in [1,d-1]$.
    This implies that $\boldsymbol{x}= \boldsymbol{y}$.
    Furthermore, since $p$ is the smallest prime larger than $\max\{P,q\}$, we have $p\leq 2\max\{P,q\}$.
    Then the redundancy of $\mathcal{C}_{del}$ can be determined by applying the pigeonhole principle directly, thereby completing the proof.
\end{IEEEproof}

\begin{remark}
    Theorem \ref{thm:bounded} generalizes the construction presented by Liu and Xing \cite{Liu-23-arxiv-sub}, who considered classical $(n,d)_q$-codes where the distance between the largest and smallest indices that two sequences differ is less than $n$ instead of $P$. 
\end{remark}

In the proof of Theorem \ref{thm:bounded}, when $q=2$, $d=3$, and $p=P>q$ is an integer, Equation (\ref{eq:VTk}) also holds, i.e., $b_1+b_2 \equiv 0 \pmod{P}$ and $b_1t_1+b_2t_2\equiv 0 \pmod{P}$. 
Since $b_1,b_2\in [1-q,q-1]=[-1,1]$, we have $\{b_1,b_2\} \in \{\{0,0\},\{-1,1\}\}$.
Then it follows by $1=t_1<t_2\leq P$ that $b_1t_1+b_2t_2=0$.
In this case, we have 
\begin{equation*}
  \begin{pmatrix}
     1 & 1 \\
    t_{1} & t_{2}
   \end{pmatrix}
   \begin{pmatrix}
     b_1 \\
     b_2
   \end{pmatrix}
   =
   \begin{pmatrix}
     0 \\
     0
   \end{pmatrix}.
\end{equation*}
We may compute $b_1=b_2=0$.
Therefore, when $q=2$ and $d=3$, Theorem \ref{thm:bounded} also holds if we set $p=P>q$.
This proves the following theorem.

\begin{theorem}\label{thm:bounded'}
    Let $a_{0} \in [0,2]$ and $a_1 \in[0, P-1]$, where $P$ is an integer.
    Define the code
    \begin{align*}
      \mathcal{C}_{2,3,P}'= 
      & \{\boldsymbol{x}\in  \Sigma_2^{n}: \mathrm{VT}^{(0)}(\boldsymbol{x}) \equiv a_0 \pmod{3}, \mathrm{VT}^{(1)}(\boldsymbol{x}) \equiv a_1 \pmod{P}\}.
    \end{align*}
    For any two distinct sequences $\boldsymbol{x},\boldsymbol{y} \in \mathcal{C}_{2,3,P}'$, if the distance between the smallest and largest indices that they differ is less than $P$, the Hamming distance between them is at least $3$.
    Moreover, by the pigeonhole principle, there exists a choice of parameters, such that $r(\mathcal{C}_{2,3,P}') \leq \log_2 P+ \log_2 3$.
\end{theorem}

\subsection{Construction of $2$-Read $(n,4)_q$-Codes}

In the subsequent theorem, we use the code constructed in Theorem \ref{thm:bounded} to design a $2$-read $(n,4)_q$-code with $2\log_q\log_q n+O(1)$ bits of redundancy.

\begin{theorem}\label{con:d>=4}
  Assume $d=4$ and $P= 2\lceil \log_q n+\log_q\log_q n\rceil+1$. Let $\mathcal{C}_{q,d,P}$ be the code constructed in Theorem \ref{thm:bounded}, the code $\mathcal{C}_{2,4}^{\ell,d}\triangleq \{\boldsymbol{x}\in \mathcal{ALL}(n,\frac{P-1}{2}): \mathbbm{1}(\boldsymbol{x}) \in \mathcal{C}_{q,d,P}\}$ is a $2$-read $(n,4)_q$-code.
  Moreover, by the pigeonhole principle, there exists a choice of parameters, such that $r(\mathcal{C}_{2,4}^{\ell,d}) \leq 2\log_q \log_q n+ \log_q (3(q-1)+1)+ 2\log_q 4+o(1)$.
\end{theorem}

\begin{IEEEproof}
    Assume $\mathcal{C}_{2,4}^{\ell,d}$ is not a $2$-read $(n,4)_q$-code, then there exist two distinct sequences $\boldsymbol{x},\boldsymbol{y} \in \mathcal{ALL}(n,\frac{P-1}{2})$ such that $d_H(\mathcal{R}(\boldsymbol{x}),\mathcal{R}(\boldsymbol{y}))\leq 3$, we get $1\leq d_H(\mathbbm{1}(\boldsymbol{x}),\mathbbm{1}(\boldsymbol{y})) \leq 3$ and the distance between the largest and smallest indices that $\mathbbm{1}(\boldsymbol{x})$ and $\mathbbm{1}(\boldsymbol{y})$ differ is at most $P-1$. This contradicts the conclusion of Theorem \ref{thm:bounded}.
    Therefore, $\mathcal{C}_{2,4}^{\ell,d}$ is a $2$-read $(n,4)_q$-code.
    Furthermore, the redundancy of $\mathcal{C}_{2,4}^{\ell,d}$ can be determined by applying the pigeonhole principle directly, thereby completing the proof.
\end{IEEEproof}

Combining this with Theorem \ref{thm:new-bound}, the following holds.

\begin{corollary}\label{cor:d=4,l=2,q>=3}
  For $q\geq 3$, the optimal redundancy of $2$-read $(n,4)_q$-codes ranges from $\log_q\log_q n-o(1)$ to $2\log_q \log_q n+ \log_q (3(q-1)+1)+ 2\log_q 4+o(1)$.
\end{corollary}

\subsection{Improved Construction of $2$-Read $(n,4)_2$-Codes}

For the binary alphabet $\{0,1\}$, the conclusion of Lemma \ref{lem:HD} can be strengthened when $d=3$.

\begin{lemma}
  Assume $\boldsymbol{x} \neq \boldsymbol{y} \in \Sigma_2^n$ with $d_H(\mathcal{R}(\boldsymbol{x}),\mathcal{R}(\boldsymbol{y}))\leq 3$, then $ d_H(\mathbbm{1}(\boldsymbol{x}),\mathbbm{1}(\boldsymbol{y}))=2$.
\end{lemma}

\begin{IEEEproof}
    Since $\boldsymbol{x}\neq \boldsymbol{y}$ and $d_H(\mathcal{R}(\boldsymbol{x}),\mathcal{R}(\boldsymbol{y}))\leq 3$, we have $d_H(\mathcal{R}(\boldsymbol{x}),\mathcal{R}(\boldsymbol{y}))\in \{2,3\}$.
    \begin{itemize}
      \item When $d_H(\mathcal{R}(\boldsymbol{x}),\mathcal{R}(\boldsymbol{y}))= 2$, by Theorem \ref{thm:1sub}, there exist two sequences $\boldsymbol{u}, \boldsymbol{v}\in \Sigma_2^{\geq 0}$, two distinct symbols $a,b \in \Sigma_2$, and an integer $t \geq 1$, such that 
            \begin{equation*}
                \begin{cases}
                  \boldsymbol{x}= (\boldsymbol{u}, \boldsymbol{\alpha}_t(ab), \boldsymbol{v}); \\
                  \boldsymbol{y}= (\boldsymbol{u}, \boldsymbol{\alpha}_t(ba), \boldsymbol{v}).
                \end{cases}
              \end{equation*}
              Let $|\boldsymbol{u}|=k$, it follows by the definition of indicator sequences that $\mathbbm{1}(x)[k+1] \neq \mathbbm{1}(y)[k+1]$, $\mathbbm{1}(x)[k+t+1] \neq \mathbbm{1}(y)[k+t+1]$, and $\mathbbm{1}(x)[i]= \mathbbm{1}(y)[i]$ for any $i \not\in \{k+1,k+t+1\}$.
              This implies that $d_H(\mathbbm{1}(\boldsymbol{x}),\mathbbm{1}(\boldsymbol{y}))=2$.
      \item When $d_H(\mathcal{R}(\boldsymbol{x}),\mathcal{R}(\boldsymbol{y}))= 3$, by Theorem \ref{thm:d=34}, there exist sequences $\boldsymbol{u}, \boldsymbol{v} \in \Sigma_2^{\geq 0}$, integers $t_1,t_2 \geq 1$, and symbols $a_1, b_1, a_2, b_2 \in \Sigma_2$ with $a_1\neq b_1$, $a_2\neq b_2$, such that $\{\{\alpha_{t_1}(a_1b_1)[t_1],a_2\}\} \neq \{\{\alpha_{t_1}(b_1a_1)[t_1],b_2\}\}$ and the following holds:
        \begin{equation*}
          \begin{cases}
            \boldsymbol{x}= (\boldsymbol{u}, \boldsymbol{\alpha}_{t_1}(a_1b_1), \boldsymbol{\alpha}_{t_2}(a_2b_2), \boldsymbol{v}); \\
            \boldsymbol{y}= (\boldsymbol{u}, \boldsymbol{\alpha}_{t_1}(b_1a_1), \boldsymbol{\alpha}_{t_2}(b_2a_2), \boldsymbol{v}).
          \end{cases}
        \end{equation*}
        Let $|\boldsymbol{u}|=k$, it follows by the definition of indicator sequences that $\mathbbm{1}(x)[k+1] \neq \mathbbm{1}(y)[k+1]$, $\mathbbm{1}(x)[k+t_1+t_2+1] \neq \mathbbm{1}(y)[k+t_1+t_2+1]$, and $\mathbbm{1}(x)[i]= \mathbbm{1}(y)[i]$ for any $i \not\in \{k+1,k+t_1+1,k+t_1+t_2+1\}$.
         Moreover, since the alphabet size is two, we have $b=1-a$ when $a\neq b \in \{0,1\}$.
         Then it follows by $a_1\neq b_1 \in \Sigma_2$ and $a_2\neq b_2 \in \Sigma_2$ that $\alpha_{t_1}(a_1b_1)[t_1]+a_2= 1-\alpha_{t_1}(b_1a_1)[t_1]+1-b_2\equiv \alpha_{t_1}(b_1a_1)[t_1]+b_2 \pmod{2}$, i.e., $\mathbbm{1}(x)[k+t_1+1]= \mathbbm{1}(y)[k+t_1+1]$.
         This implies that $d_H(\mathbbm{1}(\boldsymbol{x}),\mathbbm{1}(\boldsymbol{y}))=2$.
    \end{itemize}
    In both cases, we have $d_H(\mathbbm{1}(\boldsymbol{x}),\mathbbm{1}(\boldsymbol{y}))=2$, thereby completing the proof.
\end{IEEEproof}

This conclusion allows us to use the code $\mathcal{C}_{2,3,P}'$ to construct a $2$-read $(n,4)_2$-code, rather than using the higher redundancy code $\mathcal{C}_{2,4,P}$ directly.

\begin{theorem}\label{con:d>=4'}
  Assume $d=4$ and $P= 2\lceil \log_q n+\log_q\log_q n\rceil+1$. Let $\mathcal{C}_{2,3,P}'$ be the code constructed in Theorem \ref{thm:bounded'}, the code $\mathcal{C}\triangleq \{\boldsymbol{x}\in \mathcal{ALL}(n,\frac{P-1}{2}): \mathbbm{1}(\boldsymbol{x}) \in \mathcal{C}_{2,3,P}'\}$ is a $2$-read $(n,4)_2$-code.
  Moreover, by the pigeonhole principle, there exists a choice of parameters, such that $r(\mathcal{C}) \leq \log_2 \log_2 n+ \log_2 6+o(1)$.
\end{theorem}

The proof is similar to Theorem \ref{con:d>=4} and thus we omit it.

Combining this with Theorem \ref{thm:new-bound}, the following holds.

\begin{corollary}\label{cor:d=4,l=2,q=2}
  The optimal redundancy of $2$-read $(n,4)_2$-codes ranges from $\log_2\log_2 n-o(1)$ to $\log_2 \log_2 n+ \log_2 6+o(1)$.
\end{corollary}

\section{$2$-read codes with a minimum distance of five}\label{sec:d=5}

In this section, we explore the optimal redundancy of $2$-read $(n,5)_q$-codes.
We begin by establishing a lower bound on the redundancy.

\subsection{Lower Bound on $2$-Read Codes with a Minimum Distance of Five}

Our main contribution in this subsection is the following lemma.

\begin{lemma}\label{lem:connection}
  Let $\mathcal{C}\subseteq \Sigma_q^n$ be a $2$-read $(n,5)_q$-code. Then $\mathcal{C}$ is also a classical $(n,3)_q$-code.  
\end{lemma}

\begin{IEEEproof}
    Assume $\mathcal{C}$ is not a classical $(n,3)_q$-code, then there exist two distinct sequences  $\boldsymbol{x}$ and $\boldsymbol{y}$ in $\mathcal{C}$ such that $d_H(\boldsymbol{x},\boldsymbol{y})<3$.
    Since $\boldsymbol{x}\neq\boldsymbol{y}$, we have $d_H(\mathcal{R}(\boldsymbol{x}),\mathcal{R}(\boldsymbol{y}))\geq 2$.
    \begin{itemize}
      \item If $d_H(\boldsymbol{x},\boldsymbol{y})= 1$, it follows by the proof of Lemma  \ref{lem:HD>=2} that $d_H(\mathcal{R}(\boldsymbol{x}),\mathcal{R}(\boldsymbol{y}))= 2$.
      \item If $d_H(\boldsymbol{x},\boldsymbol{y})= 2$, let $j_1,j_2$, with $j_1<j_2$, be the indices that $\boldsymbol{x}$ and $\boldsymbol{y}$ differ.
        \begin{itemize}
          \item When $j_2=j_1+1$, it follows by the definition of $2$-read vectors that $\mathcal{R}(x)[i]= \mathcal{R}(y)[i]$ for $i \not\in [j_1,j_1+2]$. In this case, we have $d_H(\mathcal{R}(\boldsymbol{x}),\mathcal{R}(\boldsymbol{y}))\leq 3$.
          \item When $j_2 \geq j_1+2$, it follows by the definition of $2$-read vectors that  $\mathcal{R}(x)[i]= \mathcal{R}(y)[i]$ for $i \not\in \{j_1,j_1+1,j_2,j_2+1\}$. In this case, we have $d_H(\mathcal{R}(\boldsymbol{x}),\mathcal{R}(\boldsymbol{y}))\leq 4$.
        \end{itemize}
    \end{itemize}
    In all cases, we have $d_H(\mathcal{R}(\boldsymbol{x}),\mathcal{R}(\boldsymbol{y}))\leq 4$.
    This contradicts the condition that $\mathcal{C}$ is a $2$-read $(n,5)_q$-code.
    Therefore, $\mathcal{C}$ is a classical $(n,3)_q$-code.  
\end{IEEEproof}

Lemma \ref{lem:connection} allows us to leverage known results about classical error-correcting codes.
In particular, by the Hamming bound, the optimal redundancy of classical $(n,3)_q$-codes is lower bounded by $\log_q ((q-1)n+1)$. Combining this bound with Lemma \ref{lem:connection}, the following holds.

\begin{theorem}\label{thm:d=5}
  The optimal redundancy of $2$-read $(n,5)_q$-codes is lower bounded by $\log_q ((q-1)n+1)$.
\end{theorem}

\subsection{Upper Bound on $2$-Read Codes with a Minimum Distance of Five}

In this subsection, we aim to establish an upper bound on the redundancy of $2$-read $(n,5)_q$-codes by constructing such a code. 
To do so, we first present two auxiliary codes. 
The first auxiliary code forbids scenarios where the Hamming distance is less than four, as well as one specific case where the Hamming distance is equal to four. 
Meanwhile, the second auxiliary code forbids the remaining case where the Hamming distance is four.
By taking the intersection of these two auxiliary codes, we are able to construct the desired $2$-read $(n,5)_q$-code. 
In the rest of this subsection, Case (A) and Case (B) are referred to the cases outlined in Theorem \ref{thm:d=34}.

\subsubsection{The First Auxiliary Code}
We begin with an important observation. 
Assume $\boldsymbol{x}\neq \boldsymbol{y} \in \Sigma_q^n$ such that $d_H(\mathcal{R}(\boldsymbol{x}),\mathcal{R}(\boldsymbol{y}))\leq 4$. 
By Lemma \ref{lem:HD}, it follows that $1\leq d_H(\mathbbm{1}(\boldsymbol{x}),\mathbbm{1}(\boldsymbol{y}))\leq 4$.
Furthermore, if $\boldsymbol{x},\boldsymbol{y}\in \mathcal{ALL}(n,\frac{P-1}{3})$, where $P>0$ is an integer, it can be easily verified, based on Theorems \ref{thm:1sub} and \ref{thm:d=34}, that the distance between the largest and smallest indices that $\mathbbm{1}(\boldsymbol{x})$ and $\mathbbm{1}(\boldsymbol{y})$ differ is at most $\max\{t,t_1+t_2,t_1+t_2+t_3\} \leq P-1$, under the conditions that either $d_H(\mathcal{R}(\boldsymbol{x}),\mathcal{R}(\boldsymbol{y}))\leq 3$ or Case (B) occurs when $d_H(\mathcal{R}(\boldsymbol{x}),\mathcal{R}(\boldsymbol{y}))=4$.

Based on this observation, we can design a code (similar to Theorem \ref{con:d>=4}) using the construction from Theorem \ref{thm:bounded} to forbid scenarios where $d_H(\mathcal{R}(\boldsymbol{x}),\mathcal{R}(\boldsymbol{y}))\leq 3$, as well as Case (B) when $d_H(\mathcal{R}(\boldsymbol{x}),\mathcal{R}(\boldsymbol{y}))=4$.

\begin{theorem}\label{thm:aux1}
  Assume $d=5$ and $P= 3\lceil \log_q n+\log_q\log_q n\rceil+1$. Let $\mathcal{C}_{q,d,P}$ be the code constructed in Theorem \ref{thm:bounded}, we define the first auxiliary code as $\mathcal{C}_{Aux}^1\triangleq \{\boldsymbol{x}\in \mathcal{ALL}(n,\frac{P-1}{3}): \mathbbm{1}(\boldsymbol{x}) \in \mathcal{C}_{q,d,P}\}$.
  Then for any $\boldsymbol{x}\neq \boldsymbol{y} \in \mathcal{C}_{Aux}^1$, we have $d_H(\mathcal{R}(\boldsymbol{x}),\mathcal{R}(\boldsymbol{y}))\geq 4$ and Case (B) does not occur when $d_H(\mathcal{R}(\boldsymbol{x}),\mathcal{R}(\boldsymbol{y}))=4$.
  Moreover, by the pigeonhole principle, there exists a choice of parameters, such that $r(\mathcal{C}_{Aux}^1) \leq 3\log_q \log_q n+ O(1)$.
\end{theorem}

\begin{IEEEproof}
    Assume either $d_H(\mathcal{R}(\boldsymbol{x}),\mathcal{R}(\boldsymbol{y}))\leq 3$ or Case (B) occurs when $d_H(\mathcal{R}(\boldsymbol{x}),\mathcal{R}(\boldsymbol{y}))=4$, then by the previous observation, we have $1\leq d_H(\mathbbm{1}(\boldsymbol{x}),\mathbbm{1}(\boldsymbol{y})) \leq 4$ and the distance between the largest and smallest indices that $\mathbbm{1}(\boldsymbol{x})$ and $\mathbbm{1}(\boldsymbol{y})$ differ is at most $P-1$. This contradicts the conclusion of Theorem \ref{thm:bounded}, implying that $d_H(\mathcal{R}(\boldsymbol{x}),\mathcal{R}(\boldsymbol{y}))\geq 4$ and Case (B) does not occur when $d_H(\mathcal{R}(\boldsymbol{x}),\mathcal{R}(\boldsymbol{y}))= 4$.
    Furthermore, the redundancy of $\mathcal{C}_{Aux}^{1}$ can be determined by applying the pigeonhole principle directly, thereby completing the proof.
\end{IEEEproof}

\subsubsection{The Second Auxiliary Code}
We now turn our attention to designing a code that disallows the remaining scenario where the Hamming distance between $2$-read vectors is four and the conditions outlined in Case (A) are satisfied.

\begin{theorem}\label{thm:aux2}
  Let $P= 3\lceil \log_q n+\log_q\log_q n \rceil+1$ and let $p$ be the smallest prime number satisfying $p>4n$.
  Let $d_1 \in [0,2q-2], d_2 \in [0,p-1], d_3 \in [0,\lfloor \frac{(q-1)(P-1)}{3} \rfloor]$, we define the code $\mathcal{C}_{Aux}^2\subseteq \mathcal{ALL}(n,\frac{P-1}{3})$ in which each sequence satisfies the following conditions: 
  \begin{itemize}
    \item $\mathrm{VT}^{(0)}(\boldsymbol{x}) \equiv d_1 \pmod{2q-1}$ and $\mathrm{VT}^{(2)}(\boldsymbol{x}) \equiv d_2 \pmod{p}$;
    \item $\mathrm{VT}^{(0)}(\mathcal{O}(\boldsymbol{x})) \equiv d_3 \pmod{\lfloor \frac{(q-1)(P-1)}{3} +1\rfloor}$.
  \end{itemize}
  Then for any $\boldsymbol{x}\neq \boldsymbol{y} \in \mathcal{C}_{Aux}^2$, Case (A) does not occur when $d_H(\mathcal{R}(\boldsymbol{x}),\mathcal{R}(\boldsymbol{y}))=4$.
  Moreover, by the pigeonhole principle, there exists a choice of parameters, such that $r(\mathcal{C}_{Aux}^2) \leq \log_q n+ \log_q \log_q n+ O(1)$.
\end{theorem}

\begin{IEEEproof}
    Assume $d_H(\mathcal{R}(\boldsymbol{x}),\mathcal{R}(\boldsymbol{y}))= 4$ and Case (A) occurs, then by Theorem \ref{thm:d=34} there exist three sequences $\boldsymbol{u}, \boldsymbol{v}, \boldsymbol{w} \in \Sigma_q^{\geq 0}$, four symbols $a_1,b_1, a_2,b_2 \in \Sigma_q$ with $a_1 \neq b_1$ and $a_2 \neq b_2$, and two integers $t_1,t_2 \geq 1$, such that
        \begin{equation*}
          \begin{cases}
            \boldsymbol{x}= (\boldsymbol{u}, \boldsymbol{\alpha}_{t_1}(a_1b_1), \boldsymbol{v}, \boldsymbol{\alpha}_{t_2}(a_2b_2), \boldsymbol{w}); \\
            \boldsymbol{y}= (\boldsymbol{u}, \boldsymbol{\alpha}_{t_1}(b_1a_1), \boldsymbol{v}, \boldsymbol{\alpha}_{t_2}(b_2a_2), \boldsymbol{w}).
          \end{cases}
        \end{equation*}
        Set $|\boldsymbol{u}|\triangleq u$ and $|\boldsymbol{v}|\triangleq v$.
        Below we distinguish between two cases based on the parity of $t_1$.
        
        We first consider the case where $2|t_1$.
        In this case, we have $2|t_2$ since $\mathrm{VT}^{(0)}(\boldsymbol{x})- \mathrm{VT}^{(0)}(\boldsymbol{y})= a_2-b_2 \not\equiv 0 \pmod{2q-1}$ when $2\nmid t_2$.
        Then we may compute 
        \begin{equation*}
        \begin{aligned}
                  \mathrm{VT}^{(2)}(\boldsymbol{x})- \mathrm{VT}^{(2)}(\boldsymbol{y})
                  &= \sum_{j=1}^{t_1/2} ((u+2j-1)^2(a_1-b_1)+ (u+2j)^2(b_1-a_1)) \\
                  & ~~~~+ \sum_{j=1}^{t_2/2} ((u+t_1+v+2j-1)^2(a_2-b_2)+ (u+t_1+v+2j)^2(b_2-a_2)) \\
                  &= \sum_{j=1}^{t_1/2} (2u+4j-1)(b_1-a_1)+ \sum_{j=1}^{t_1/2} (2u+2t_1+2v+4j-1)(b_2-a_2)\\
                  &= \frac{t_1}{2}(2u+t_1+1)(b_1-a_1)+ \frac{t_2}{2}(2u+2t_1+2v+t_2+1)(b_2-a_2).
        \end{aligned}
        \end{equation*}
        Below we further calculate $\mathrm{VT}^{(2)}(\boldsymbol{x})- \mathrm{VT}^{(2)}(\boldsymbol{y})$ by considering the parity of $u$ and $v$ and using the condition that $\mathrm{VT}^{(0)}(\mathcal{O}(\boldsymbol{x}))\equiv \mathrm{VT}^{(0)}(\mathcal{O}(\boldsymbol{y})) \pmod{\lfloor \frac{(q-1)P}{3}+1 \rfloor}$.
        \begin{itemize}
          \item If $2|u$ and $2|v$, we have
                \begin{equation*}
                  \begin{cases}
                    \mathcal{O}(\boldsymbol{x})= (\mathcal{O}(\boldsymbol{u}), {a_1}^{t_1/2}, \mathcal{O}(\boldsymbol{v}), {a_2}^{t_2/2}, \mathcal{O}(\boldsymbol{w})); \\
                    \mathcal{O}(\boldsymbol{y})= (\mathcal{O}(\boldsymbol{u}), {b_1}^{t_1/2}, \mathcal{O}(\boldsymbol{v}), {b_2}^{t_2/2}, \mathcal{O}(\boldsymbol{w})).
                  \end{cases}
                \end{equation*}
                We may compute $|\mathrm{VT}^{(0)}(\mathcal{O}(\boldsymbol{x}))- \mathrm{VT}^{(0)}(\mathcal{O}(\boldsymbol{y}))|= |\frac{t_1}{2}(a_1-b_1)+ \frac{t_2}{2}(a_2-b_2)|\leq \lfloor \frac{(q-1)(P-1)}{3}\rfloor$, then it follows by $\mathrm{VT}^{(0)}(\mathcal{O}(\boldsymbol{x}))\equiv \mathrm{VT}^{(0)}(\mathcal{O}(\boldsymbol{y})) \pmod{\lfloor \frac{(q-1)(P-1)}{3} +1\rfloor}$ that $t_1(a_1-b_1)=t_2(b_2-a_2)$.
                In this case, we may further compute $\mathrm{VT}^{(2)}(\boldsymbol{x})- \mathrm{VT}^{(2)}(\boldsymbol{y})$ as follows:
                \begin{equation}\label{eq:VT_even'}
                \begin{aligned}
                  \mathrm{VT}^{(2)}(\boldsymbol{x})- \mathrm{VT}^{(2)}(\boldsymbol{y})
                  &= \frac{t_1}{2}(2u+t_1+1)(b_1-a_1)+ \frac{t_2}{2}(2u+2t_1+2v+t_2+1)(b_2-a_2) \\
                  &= \frac{t_1}{2}(a_1-b_1)(-2u-t_1-1+2u+2t_1+2v+t_2+1) \\
                  &= \frac{t_1}{2}(a_1-b_1)(t_1+2v+t_2).
                \end{aligned}
                \end{equation}
                Since $\max\{t_1,|a_1-b_1|,t_1+2v+t_2\}<4n$ and $p$ is a prime larger than $4n$, we get $\mathrm{VT}^{(2)}(\boldsymbol{x})- \mathrm{VT}^{(2)}(\boldsymbol{y})\not\equiv 0 \pmod{p}$, which leads to a contradiction.
          \item If $2 \nmid u$ and $2 | v$, we have
                \begin{equation*}
                  \begin{cases}
                    \mathcal{O}(\boldsymbol{x})= (\mathcal{O}(\boldsymbol{u}), {b_1}^{t_1/2}, \mathcal{E}(\boldsymbol{v}), {b_2}^{t_2/2}, \mathcal{E}(\boldsymbol{w})); \\
                    \mathcal{O}(\boldsymbol{y})= (\mathcal{O}(\boldsymbol{u}), {a_1}^{t_1/2}, \mathcal{E}(\boldsymbol{v}), {a_2}^{t_2/2}, \mathcal{E}(\boldsymbol{w})).
                  \end{cases}
                \end{equation*}
                We may compute $|\mathrm{VT}^{(0)}(\mathcal{O}(\boldsymbol{x}))- \mathrm{VT}^{(0)}(\mathcal{O}(\boldsymbol{y}))|= |\frac{t_1}{2}(b_1-a_1)+ \frac{t_2}{2}(b_2-a_2)|\leq \lfloor \frac{(q-1)(P-1)}{3}\rfloor$, then it follows by $\mathrm{VT}^{(0)}(\mathcal{O}(\boldsymbol{x}))\equiv \mathrm{VT}^{(0)}(\mathcal{O}(\boldsymbol{y})) \pmod{\lfloor \frac{(q-1)(P-1)}{3} +1\rfloor}$ that $t_1(a_1-b_1)=t_2(b_2-a_2)$.
                In this case, Equation (\ref{eq:VT_even'}) holds, which leads to a contradiction.
          \item If $2|u$ and $2 \nmid v$, we have
                \begin{equation*}
                  \begin{cases}
                    \mathcal{O}(\boldsymbol{x})= (\mathcal{O}(\boldsymbol{u}), {a_1}^{t_1/2}, \mathcal{O}(\boldsymbol{v}), {b_2}^{t_2/2}, \mathcal{E}(\boldsymbol{w})); \\
                    \mathcal{O}(\boldsymbol{y})= (\mathcal{O}(\boldsymbol{u}), {b_1}^{t_1/2}, \mathcal{O}(\boldsymbol{v}), {a_2}^{t_2/2}, \mathcal{E}(\boldsymbol{w})).
                  \end{cases}
                \end{equation*}
                We may compute $|\mathrm{VT}^{(0)}(\mathcal{O}(\boldsymbol{x}))- \mathrm{VT}^{(0)}(\mathcal{O}(\boldsymbol{y}))|= |\frac{t_1}{2}(a_1-b_1)+ \frac{t_2}{2}(b_2-a_2)|\leq \lfloor \frac{(q-1)(P-1)}{3}\rfloor$, then it follows by $\mathrm{VT}^{(0)}(\mathcal{O}(\boldsymbol{x}))\equiv \mathrm{VT}^{(0)}(\mathcal{O}(\boldsymbol{y})) \pmod{\lfloor \frac{(q-1)(P-1)}{3} +1\rfloor}$ that $t_1(b_1-a_1)=t_2(b_2-a_2)$.
                In this case, we may further compute $\mathrm{VT}^{(2)}(\boldsymbol{x})- \mathrm{VT}^{(2)}(\boldsymbol{y})$ as follows:
                \begin{equation}\label{eq:VT_even''}
                \begin{aligned}
                  \mathrm{VT}^{(2)}(\boldsymbol{x})- \mathrm{VT}^{(2)}(\boldsymbol{y})
                  &= \frac{t_1}{2}(2u+t_1+1)(b_1-a_1)+ \frac{t_2}{2}(2u+2t_1+2v+t_2+1)(b_2-a_2) \\
                  &= \frac{t_1}{2}(b_1-a_1)(2u+t_1+1+2u+2t_1+2v+t_2+1) \\
                  &= \frac{t_1}{2}(b_1-a_1)(4u+3t_1+2v+t_2+2).
                \end{aligned}
                \end{equation}
                Since $\max\{t_1,|a_1-b_1|,4u+3t_1+2v+t_2+2\}<4n$ and $p$ is a prime larger than $4n$, we get $\mathrm{VT}^{(2)}(\boldsymbol{x})- \mathrm{VT}^{(2)}(\boldsymbol{y})\not\equiv 0 \pmod{p}$, which leads to a contradiction.
          \item If $2 \nmid u$ and $2 \nmid v$, we have
                \begin{equation*}
                  \begin{cases}
                    \mathcal{O}(\boldsymbol{x})= (\mathcal{O}(\boldsymbol{u}), {b_1}^{t_1/2}, \mathcal{E}(\boldsymbol{v}), {a_2}^{t_2/2}, \mathcal{O}(\boldsymbol{w})); \\
                    \mathcal{O}(\boldsymbol{y})= (\mathcal{O}(\boldsymbol{u}), {a_1}^{t_1/2}, \mathcal{E}(\boldsymbol{v}), {b_2}^{t_2/2}, \mathcal{O}(\boldsymbol{w})).
                  \end{cases}
                \end{equation*}
                We may compute $|\mathrm{VT}^{(0)}(\mathcal{O}(\boldsymbol{x}))- \mathrm{VT}^{(0)}(\mathcal{O}(\boldsymbol{y}))|= |\frac{t_1}{2}(b_1-a_1)+ \frac{t_2}{2}(a_2-b_2)|\leq \lfloor \frac{(q-1)(P-1)}{3}\rfloor$, then it follows by $\mathrm{VT}^{(0)}(\mathcal{O}(\boldsymbol{x}))\equiv \mathrm{VT}^{(0)}(\mathcal{O}(\boldsymbol{y})) \pmod{\lfloor \frac{(q-1)(P-1)}{3} +1\rfloor}$ that $t_1(b_1-a_1)=t_2(b_2-a_2)$.
                In this case, Equation (\ref{eq:VT_even''}) holds, which leads to a contradiction.
        \end{itemize}
        
        We now consider the case where $2 \nmid t_1$.
        In this case, we have $2 \nmid t_2$ since $\mathrm{VT}^{(0)}(\boldsymbol{x})- \mathrm{VT}^{(0)}(\boldsymbol{y})= a_1-b_1 \not\equiv 0 \pmod{2q-1}$ when $2|t_2$.
        Then we have $a_1+a_2 \equiv b_1+b_2 \pmod{2q-1}$ since $\mathrm{VT}^{(0)}(\boldsymbol{x})- \mathrm{VT}^{(0)}(\boldsymbol{y})= a_1+a_2-b_1-b_2 \equiv 0 \pmod{2q-1}$.
        It follows by $a_1,b_1,a_2,b_2 \in \Sigma_q$ that $a_1+a_2=b_1+b_2$, i.e., $a_1-b_1=b_2-a_2$.
        We may compute
        \begin{equation*}
        \begin{aligned}
                  \mathrm{VT}^{(2)}(\boldsymbol{x})- \mathrm{VT}^{(2)}(\boldsymbol{y})
                  &= \sum_{j=1}^{(t_1-1)/2} ((u+2j-1)^2(a_1-b_1)+ (u+2j)^2(b_1-a_1))\\
                  & ~~~~+ \sum_{j=1}^{(t_2-1)/2} ((u+t_1+v+2j-1)^2(a_2-b_2)+ (u+t_1+v+2j)^2(b_2-a_2)) \\
                  & ~~~~+ (u+t_1)^2(a_1-b_1) + (u+t_1+v+t_2)^2(a_2-b_2) \\
                  &= \sum_{j=1}^{(t_1-1)/2} (2u+4j-1)(a_2-b_2)+ \sum_{j=1}^{(t_2-1)/2} (2u+2t_1+2v+4j-1)(b_2-a_2) \\
                  &~~~~+ (v+t_2)(2u+2t_1+v+t_2)(a_2-b_2).
                \end{aligned}
        \end{equation*}
        Below we further calculate $\mathrm{VT}^{(2)}(\boldsymbol{x})- \mathrm{VT}^{(2)}(\boldsymbol{y})$ by considering the parity of $u$ and $v$ and using the condition that $\mathrm{VT}^{(0)}(\mathcal{O}(\boldsymbol{x}))\equiv \mathrm{VT}^{(0)}(\mathcal{O}(\boldsymbol{y})) \pmod{\lfloor \frac{(q-1)(P-1)}{3} +1\rfloor}$.
        \begin{itemize}
          \item If $2|u$ and $2|v$, we have
                \begin{equation*}
                  \begin{cases}
                    \mathcal{O}(\boldsymbol{x})= (\mathcal{O}(\boldsymbol{u}), {a_1}^{(t_1+1)/2}, \mathcal{E}(\boldsymbol{v}), {b_2}^{(t_2-1)/2}, \mathcal{O}(\boldsymbol{w})); \\
                    \mathcal{O}(\boldsymbol{y})= (\mathcal{O}(\boldsymbol{u}), {b_1}^{(t_1+1)/2}, \mathcal{E}(\boldsymbol{v}), {a_2}^{(t_2-1)/2}, \mathcal{O}(\boldsymbol{w})).
                  \end{cases}
                \end{equation*}
                We may compute $|\mathrm{VT}^{(0)}(\mathcal{O}(\boldsymbol{x}))- \mathrm{VT}^{(0)}(\mathcal{O}(\boldsymbol{y}))|= |\frac{t_1+1}{2}(a_1-b_1)+ \frac{t_2-1}{2}(b_2-a_2)|= |a_1-b_1| \frac{t_1+t_2}{2}$.
                Note that $0<|a_1-b_1| \frac{t_1+t_2}{2} \leq \lfloor \frac{(q-1)(P-1)}{3} \rfloor$, this contradicts the condition that $\mathrm{VT}^{(0)}(\mathcal{O}(\boldsymbol{x}))\equiv \mathrm{VT}^{(0)}(\mathcal{O}(\boldsymbol{y})) \pmod{\lfloor \frac{(q-1)(P-1)}{3} +1\rfloor}$.
                Therefore, this case will not occur.
          \item If $2 \nmid u$ and $2 | v$, we have
                \begin{equation*}
                  \begin{cases}
                    \mathcal{O}(\boldsymbol{x})= (\mathcal{O}(\boldsymbol{u}), {b_1}^{(t_1-1)/2}, \mathcal{O}(\boldsymbol{v}), {a_2}^{(t_2+1)/2}, \mathcal{E}(\boldsymbol{w})); \\
                    \mathcal{O}(\boldsymbol{y})= (\mathcal{O}(\boldsymbol{u}), {a_1}^{(t_1-1)/2}, \mathcal{O}(\boldsymbol{v}), {b_2}^{(t_2+1)/2}, \mathcal{E}(\boldsymbol{w})).
                  \end{cases}
                \end{equation*}
                Similar to the previous case, we may compute $\mathrm{VT}^{(0)}(\mathcal{O}(\boldsymbol{x}))- \mathrm{VT}^{(0)}(\mathcal{O}(\boldsymbol{y}))= \frac{t_1-1}{2}(b_1-a_1)+ \frac{t_2+1}{2}(a_2-b_2)= (b_1-a_1) \frac{t_1+t_2}{2} \not\equiv 0 \pmod{\lfloor \frac{(q-1)(P-1)}{3} +1\rfloor}$, which leads to a contradiction.
                Therefore, this case will not occur.
          \item If $2|u$ and $2 \nmid v$, we have
                \begin{equation*}
                  \begin{cases}
                    \mathcal{O}(\boldsymbol{x})= (\mathcal{O}(\boldsymbol{u}), {a_1}^{(t_1+1)/2}, \mathcal{E}(\boldsymbol{v}), {a_2}^{(t_2+1)/2}, \mathcal{E}(\boldsymbol{w})); \\
                    \mathcal{O}(\boldsymbol{y})= (\mathcal{O}(\boldsymbol{u}), {b_1}^{(t_1+1)/2}, \mathcal{E}(\boldsymbol{v}), {b_2}^{(t_2+1)/2}, \mathcal{E}(\boldsymbol{w})).
                  \end{cases}
                \end{equation*}
                We may compute $|\mathrm{VT}^{(0)}(\mathcal{O}(\boldsymbol{x}))- \mathrm{VT}^{(0)}(\mathcal{O}(\boldsymbol{y}))|= |\frac{t_1+1}{2}(a_1-b_1)+ \frac{t_2+1}{2}(a_2-b_2)|= |(a_1-b_1) \frac{t_1-t_2}{2}|\leq \lfloor \frac{(q-1)(P-1)}{3}\rfloor$, then it follows by $\mathrm{VT}^{(0)}(\mathcal{O}(\boldsymbol{x}))\equiv \mathrm{VT}^{(0)}(\mathcal{O}(\boldsymbol{y})) \pmod{\lfloor \frac{(q-1)(P-1)}{3} +1\rfloor}$ that $t_1=t_2$.
                In this case, we may further compute $\mathrm{VT}^{(2)}(\boldsymbol{x})- \mathrm{VT}^{(2)}(\boldsymbol{y})$ as follows:
                \begin{equation}\label{eq:VT_odd'}
                \begin{aligned}
                  \mathrm{VT}^{(2)}(\boldsymbol{x})- \mathrm{VT}^{(2)}(\boldsymbol{y})
                  &= \sum_{j=1}^{(t_1-1)/2} (2u+4j-1)(a_2-b_2)+ \sum_{j=1}^{(t_2-1)/2} (2u+2t_1+2v+4j-1)(b_2-a_2) \\
                  &~~~~+ (v+t_2)(2u+2t_1+v+t_2)(a_2-b_2)\\
                  &= \sum_{j=1}^{(t_1-1)/2} (2v+2t_1)(b_2-a_2)+ (v+t_1)(2u+3t_1+v)(a_2-b_2)\\
                  &= (t_1-1)(v+t_1)(b_2-a_2)+ (v+t_1)(2u+3t_1+v)(a_2-b_2) \\
                  &= (a_2-b_2)(v+t_1)(2u+v+2t_1+1).
                \end{aligned}
                \end{equation}
                Since $\max\{|a_2-b_2|,v+t_1,2u+v+2t_1+1\}<4n$ and $p$ is a prime larger than $4n$, we get $\mathrm{VT}^{(2)}(\boldsymbol{x})- \mathrm{VT}^{(2)}(\boldsymbol{y})\not\equiv 0 \pmod{p}$, which leads to a contradiction.
          \item If $2 \nmid u$ and $2 \nmid v$, we have
                \begin{equation*}
                  \begin{cases}
                    \mathcal{O}(\boldsymbol{x})= (\mathcal{O}(\boldsymbol{u}), {b_1}^{(t_1-1)/2}, \mathcal{O}(\boldsymbol{v}), {b_2}^{(t_2-1)/2}, \mathcal{O}(\boldsymbol{w})); \\
                    \mathcal{O}(\boldsymbol{y})= (\mathcal{O}(\boldsymbol{u}), {a_1}^{(t_1-1)/2}, \mathcal{O}(\boldsymbol{v}), {a_2}^{(t_2-1)/2}, \mathcal{O}(\boldsymbol{w})).
                  \end{cases}
                \end{equation*}
                We may compute $|\mathrm{VT}^{(0)}(\mathcal{O}(\boldsymbol{x}))- \mathrm{VT}^{(0)}(\mathcal{O}(\boldsymbol{y}))|= |\frac{t_1-1}{2}(b_1-a_1)+ \frac{t_2-1}{2}(b_2-a_2)|= |(b_1-a_1) \frac{t_1-t_2}{2}|\leq \lfloor \frac{(q-1)(P-1)}{3}\rfloor$, then it follows by $\mathrm{VT}^{(0)}(\mathcal{O}(\boldsymbol{x}))\equiv \mathrm{VT}^{(0)}(\mathcal{O}(\boldsymbol{y})) \pmod{\lfloor \frac{(q-1)(P-1)}{3} +1\rfloor}$ that $t_1=t_2$.
                In this case, Equation (\ref{eq:VT_odd'}) holds, which leads to a contradiction.
        \end{itemize}
        
        In all cases, we obtain a contradiction, implying that Case (A) does not occur when $d_H(\mathcal{R}(\boldsymbol{x}),\mathcal{R}(\boldsymbol{y}))= 4$.
        Furthermore, the redundancy of $\mathcal{C}_{Aux}^{2}$ can be determined by applying the pigeonhole principle directly, thereby completing the proof.
\end{IEEEproof}

We can now design the desired $2$-read $(n,5)_q$-code by taking the intersection of the two auxiliary codes constructed previously in Theorems \ref{thm:aux1} and \ref{thm:aux2}. 
This leads to the following theorem.

\begin{theorem}\label{con:d>=5}
  Let $\mathcal{C}_{Aux}^1$ and $\mathcal{C}_{Aux}^2$ be the codes defined in Theorems \ref{thm:aux1} and \ref{thm:aux2}, respectively.
  Then the code $\mathcal{C}_{2,5}^{\ell,d}\triangleq \mathcal{C}_{Aux}^1 \cap \mathcal{C}_{Aux}^2$ is a $2$-read $(n,5)_q$-code.
  Moreover, by the pigeonhole principle, there exists a choice of parameters, such that $r(\mathcal{C}_{2,5}^{\ell,d}) \leq \log_q n+ 4\log_q \log_q n+ O(1)$.
\end{theorem}

Combining this with Theorem \ref{thm:d=5}, the following holds.
\begin{corollary}\label{cor:d=5,l=2}
  The optimal redundancy of $2$-read $(n,5)_q$-codes ranges from $\log_q n+O(1)$ to $\log_q n+ 4\log_q\log_q n+O(1)$.
\end{corollary}

\begin{remark}
  For two sequences $\boldsymbol{x}$ and $\boldsymbol{y}$ in $\Sigma_2^n$ such that $\mathcal{D}_1(\boldsymbol{x})\cap \mathcal{D}_1(\boldsymbol{y})= \emptyset$ and $|\mathcal{D}_2(\boldsymbol{x})\cap \mathcal{D}_2(\boldsymbol{y})|\geq 4$, 
  we characterized the structure of them in \cite{Sun-23-IT-DR}. 
  One of the main cases corresponds to the scenario prohibited in the second auxiliary code discussed in this subsection. 
  This suggests a strong connection between $2$-read $(n,5)_q$-codes and classical $(n,4;\mathcal{D}_2)$-reconstruction codes. 
  However, our $2$-read $(n,5)_q$-code may not qualify as a classical $(n,4;\mathcal{D}_2)$-reconstruction code, as the condition $\mathcal{D}_1(\boldsymbol{x})\cap \mathcal{D}_1(\boldsymbol{y})= \emptyset$ may not be met in our $2$-read $(n,5)_q$-code.
\end{remark}

\section{$\ell$-Read Codes under the Reconstruction Model}\label{sec:recon}

In this section, we consider $\ell$-read codes under the reconstruction model, where the goal is to recover the original message from multiple noisy reads. 
Levenshtein \cite{Levenshtein-01-IT-recons} proved that the minimum number of distinct noisy reads required to reconstruct the original message is one more than the largest intersection size between the error balls of any two distinct transmitted messages.
In other words, in classical codes, let $\mathcal{C}$ be a classical $(n,d)_q$-code and set $N(n,q,t,d;\mathcal{C}) \triangleq \max \{|\mathcal{S}_t(\boldsymbol{x})\cap \mathcal{S}_t(\boldsymbol{y})|:\boldsymbol{x}, \boldsymbol{y} \in \mathcal{C}\}$, then $\mathcal{C}$ is a classical $(n,N;\mathcal{S}_t)_q$-reconstruction code when $N\geq N(n,q,t,d;\mathcal{C})+1$.
Moreover, let $N(n,q,t,d)\triangleq \{|\mathcal{S}_t(\boldsymbol{x})\cap \mathcal{S}_t(\boldsymbol{y})|:\boldsymbol{x}, \boldsymbol{y} \in \Sigma_q^n, d_H(\boldsymbol{x},\boldsymbol{y})\geq d\}$, then $N(n,q,t,d)\geq N(n,q,t,d;\mathcal{C})$ for any classical $(n,d)_q$-code $\mathcal{C}$. The quantity of $N(n,q,t,d)$ is determined in \cite{Levenshtein-01-IT-recons}.

\begin{lemma}[Corollary 1 of \cite{Levenshtein-01-IT-recons}]\label{lem:recon}
    Assume $d\geq 1$ and $t \geq \lceil d/2 \rceil$, then
    \begin{align*}
      N(n,q,t,d)= \sum_{i=0}^{t-\lceil d/2 \rceil} \binom{n-d}{i}(q-1)^i \sum_{k=d-t+i}^{t-i} \sum_{l=d-t+i}^{t-i} \binom{d}{k} \binom{d-k}{l}(q-2)^{d-k-l},
    \end{align*}
    where $\binom{n}{i}\triangleq 0$ if $i>n$.
\end{lemma}

Similarly, in $\ell$-read codes, let $\mathcal{C}$ be an $\ell$-read $(n,d)_q$-code and set $N(n,\ell,q,t,d;\mathcal{C}) \triangleq \max \{|\mathcal{S}_t(\mathcal{R}_{\ell}(\boldsymbol{x}))\cap \mathcal{S}_t(\mathcal{R}_{\ell}(\boldsymbol{y}))|:\boldsymbol{x}, \boldsymbol{y} \in \mathcal{C}\}$, then $\mathcal{C}$ is an $\ell$-read $(n,N;\mathcal{S}_t)_q$-reconstruction code when $N\geq N(n,\ell,q,t,d;\mathcal{C})+1$.
Moreover, let $N(n,\ell,q,t,d) \triangleq \{|\mathcal{S}_t(\mathcal{R}_{\ell}(\boldsymbol{x}))\cap \mathcal{S}_t(\mathcal{R}_{\ell}(\boldsymbol{y}))|:\boldsymbol{x}, \boldsymbol{y} \in \Sigma_q^n,d_H(\mathcal{R}_{\ell}(\boldsymbol{x}),\mathcal{R}_{\ell}(\boldsymbol{y}))\geq d\}$, then $N(n,\ell,q,t,d)\geq N(n,\ell,q,t,d;\mathcal{C})$ for any $\ell$-read $(n,d)_q$-code $\mathcal{C}$. 
Therefore, the main task in this section is to determine $N(n,\ell,q,t,d)$ or present an upper bound of $N(n,\ell,q,t,d)$.
To do so, we first establish the connection between $\ell$-read vectors and $\binom{q+\ell-1}{\ell}$-ary sequences of length $n+\ell-1$.

\begin{definition}
  Let $\mathcal{M}_{q,\ell}$ be the set of all multi-sets composed of $\ell$ symbols from the alphabet $\Sigma_q$ and $\mathcal{M}_{q,\ell}^{n+\ell-1}$ be the set of all vectors of length $n+\ell-1$ over the set $\mathcal{M}_{q,\ell}$.
\end{definition}

It is well known that $|\mathcal{M}_{q,\ell}|= \binom{q+\ell-1}{\ell}\triangleq q_{\ell}$.
Therefore, there exists a bijection between the set $\mathcal{M}_{q,\ell}$ and the alphabet $\Sigma_{q_{\ell}}$. 
Let $\phi_{\ell}$ be such a bijection and let $\Phi_{\ell}(\boldsymbol{X})\triangleq (\phi_{\ell}(X[1]), \phi_{\ell}(X[2]),\ldots, \phi_{\ell}(X[n+\ell-1]))$ for any $\boldsymbol{X} \in \mathcal{M}_{q,\ell}^{n+\ell-1}$, then $\Phi_{\ell}$ is a bijection between $\mathcal{M}_{q,\ell}^{n+\ell-1}$ and $\Sigma_{q_{\ell}}^{n+\ell-1}$. 

We now present an upper bound of $N(n,\ell,q,t,d)$.

\begin{lemma}
  Let $q_{\ell}= \binom{q+\ell-1}{\ell}$, then $N(n,\ell,q,t,d)\leq N(n+\ell-1,q_{\ell},t,d)$.
\end{lemma}

\begin{IEEEproof}
    Assume $\boldsymbol{x},\boldsymbol{y} \in \Sigma_q^n$ with $d_H(\mathcal{R}_{\ell}(\boldsymbol{x}),\mathcal{R}_{\ell}(\boldsymbol{y}))\geq d$, we have $d_H(\Phi_{\ell}(\mathcal{R}_{\ell}(\boldsymbol{x})),\Phi_{\ell}(\mathcal{R}_{\ell}(\boldsymbol{y})))= d_H(\mathcal{R}_{\ell}(\boldsymbol{x}),\mathcal{R}_{\ell}(\boldsymbol{y})) \geq d$ and $|\mathcal{S}_t(\mathcal{R}_{\ell}(\boldsymbol{x}))\cap \mathcal{S}_t(\mathcal{R}_{\ell}(\boldsymbol{y}))|= |\mathcal{S}_t(\Phi_{\ell}(\mathcal{R}_{\ell}(\boldsymbol{x})))\cap \mathcal{S}_t(\Phi_{\ell}(\mathcal{R}_{\ell}(\boldsymbol{y})))|\leq N(n+\ell-1,q_{\ell},t,d)$.
    Then the conclusion follows.
\end{IEEEproof}

Combining this conclusion with the codes constructed in the previous three sections, we obtain the following theorem.

\begin{theorem}
   The following statements are true.
   \begin{itemize}
     \item Assume $d=5$, $t \geq 3$, $\ell=2$, $q\geq 2$, $q_{\ell}= \binom{q+\ell-1}{\ell}$, and $N \geq N(n+\ell-1,q_{\ell},t,d)+1$, there exists a $2$-read $(n,N;\mathcal{S}_t)_q$-reconstruction code with at most $\log_q n +4\log_q \log_q n+O(1)$ bits of redundancy.
     \item Assume $d=4$, $t \geq 2$, $\ell=2$, $q\geq 3$, $q_{\ell}= \binom{q+\ell-1}{\ell}$, and $N \geq N(n+\ell-1,q_{\ell},t,d)+1$, there exists a $2$-read $(n,N;\mathcal{S}_t)_q$-reconstruction code with at most $2\log_q \log_q n+O(1)$ bits of redundancy.
     \item Assume $d=4$, $t \geq 2$, $\ell=2$, $q=2$, $q_{\ell}= \binom{q+\ell-1}{\ell}$, and $N \geq N(n+\ell-1,q_{\ell},t,d)+1$, there exists a $2$-read $(n,N;\mathcal{S}_t)_q$-reconstruction code with at most $\log_q \log_q n+O(1)$ bits of redundancy.
     \item Assume $d=3$, $t \geq 2$, $\ell\geq 2$, $q\geq 2$, $q_{\ell}= \binom{q+\ell-1}{\ell}$, and $N \geq N(n+\ell-1,q_{\ell},t,d)+1$, there exists an $\ell$-read $(n,N;\mathcal{S}_t)_q$-reconstruction code with at most $\log_q \log_q n+O(1)$ bits of redundancy.
   \end{itemize}
\end{theorem}

\begin{remark}
    In Section V of \cite{Banerjee-23-arxiv-nanopore},  Banerjee \emph{et al.} showed that when $\ell\geq 3$, $\mathcal{C}\subseteq \Sigma_q^n$ is an $\ell$-read $(n,1;\mathcal{S}_1)_q$-code ($\ell$-read $(n,3)_q$-code) if and only if it is an $\ell$-read $(n,2;\mathcal{S}_1)_q$-code.
    In fact, this conclusion also holds when $\ell=2$.
    Therefore, the optimal redundancy of $\ell$-read $(n,2;\mathcal{S}_1)_q$-code asymptotically approaches to $\log_q\log_q n-\log_q 2$ when $\ell \geq 3$, and ranges from $\log_q\log_q n-o(1)$ to $\log_q\log_q n+ 1-\log_q 2+o(1)$ when $\ell=2$.
\end{remark}

\section{Conclusion}\label{sec:concl}

In this paper, we focus on the bounds and constructions of $\ell$-read $(n,d)_q$-codes.
Firstly, we present the characterization of two sequences when their $2$-read vectors have a Hamming distance of exactly $d$.
Then we investigate the bounds and constructions of $2$-read codes with a minimum distance of $3$, $4$, and $5$, respectively.
Moreover, when $d=3$, we extend our investigation to $\ell$-read codes with a general read length $\ell \geq 2$ rather than restricting to $\ell=2$.
Our results indicate that when $\ell\geq 2$, $d=3$ and $\ell=2$, $d=4$, the optimal redundancy of $\ell$-read $(n,d)_q$-codes is $o(\log_q n)$, while for $\ell=2$, $d=5$ it is $(1+o(1))\log_q n$.
Further exploration of $\ell$-read codes with a larger minimum distance and a general read length would be an intriguing direction for future research.

\end{document}